\documentclass[twocolumn,aps,prl,shownopacs,superscriptaddress]{revtex4}
\usepackage{graphicx}
\usepackage{amssymb}
\usepackage{amsmath}
\usepackage{amsthm}
\usepackage{wasysym}
\usepackage{array}
\usepackage{xcolor}
\usepackage[colorlinks=true,linkcolor=blue,urlcolor=blue,citecolor=blue,pdfauthor={ },pdftitle={},pdfsubject={ },pdfkeywords={ }]{hyperref}

\begin{document}
	
	\title{Experimental Adiabatic Quantum Metrology with the Heisenberg scaling}
	
	\date{\today}
\author{Ran Liu}
\thanks{These two authors contributed equally}
\affiliation{Hefei National Laboratory for Physical Sciences at the Microscale and Department of Modern Physics, University of Science and Technology of China, Hefei 230026, China}
\affiliation{CAS Key Laboratory of Microscale Magnetic Resonance, University of Science and Technology of China, Hefei 230026, China}
\affiliation{Synergetic Innovation Center of Quantum Information and Quantum Physics, University of Science and Technology of China, Hefei 230026, China}
\author{Yu Chen}
\thanks{These two authors contributed equally}
\affiliation{Department of Mechanical and Automation Engineering, The Chinese University of Hong Kong, Shatin, Hong Kong SAR, China}
	\author{Min Jiang}
\affiliation{Hefei National Laboratory for Physical Sciences at the Microscale and Department of Modern Physics, University of Science and Technology of China, Hefei 230026, China}
\affiliation{CAS Key Laboratory of Microscale Magnetic Resonance, University of Science and Technology of China, Hefei 230026, China}
\affiliation{Synergetic Innovation Center of Quantum Information and Quantum Physics, University of Science and Technology of China, Hefei 230026, China}
\author{Xiaodong Yang}
\affiliation{Hefei National Laboratory for Physical Sciences at the Microscale and Department of Modern Physics, University of Science and Technology of China, Hefei 230026, China}
\affiliation{CAS Key Laboratory of Microscale Magnetic Resonance, University of Science and Technology of China, Hefei 230026, China}
\affiliation{Synergetic Innovation Center of Quantum Information and Quantum Physics, University of Science and Technology of China, Hefei 230026, China}
	\author{Ze Wu}
\affiliation{Hefei National Laboratory for Physical Sciences at the Microscale and Department of Modern Physics, University of Science and Technology of China, Hefei 230026, China}
\affiliation{CAS Key Laboratory of Microscale Magnetic Resonance, University of Science and Technology of China, Hefei 230026, China}
\affiliation{Synergetic Innovation Center of Quantum Information and Quantum Physics, University of Science and Technology of China, Hefei 230026, China}
\author{Yuchen Li}
\affiliation{Hefei National Laboratory for Physical Sciences at the Microscale and Department of Modern Physics, University of Science and Technology of China, Hefei 230026, China}
\affiliation{CAS Key Laboratory of Microscale Magnetic Resonance, University of Science and Technology of China, Hefei 230026, China}
\affiliation{Synergetic Innovation Center of Quantum Information and Quantum Physics, University of Science and Technology of China, Hefei 230026, China}
\author{Haidong Yuan}
\email{hdyuan@mae.cuhk.edu.hk}
\affiliation{Department of Mechanical and Automation Engineering, The Chinese University of Hong Kong, Shatin, Hong Kong SAR, China}
\author{Xinhua Peng}
\email{xhpeng@ustc.edu.cn}
\affiliation{Hefei National Laboratory for Physical Sciences at the Microscale and Department of Modern Physics, University of Science and Technology of China, Hefei 230026, China}
\affiliation{CAS Key Laboratory of Microscale Magnetic Resonance, University of Science and Technology of China, Hefei 230026, China}
\affiliation{Synergetic Innovation Center of Quantum Information and Quantum Physics, University of Science and Technology of China, Hefei 230026, China}
\author{Jiangfeng Du}
\affiliation{Hefei National Laboratory for Physical Sciences at the Microscale and Department of Modern Physics, University of Science and Technology of China, Hefei 230026, China}
\affiliation{CAS Key Laboratory of Microscale Magnetic Resonance, University of Science and Technology of China, Hefei 230026, China}
\affiliation{Synergetic Innovation Center of Quantum Information and Quantum Physics, University of Science and Technology of China, Hefei 230026, China}
	
	\begin{abstract}
	The critical quantum metrology, which exploits the quantum phase transition for high precision measurement, has gained increasing attention recently. The critical quantum metrology with the continuous quantum phase transition, however, is experimentally very challenging since the continuous quantum phase transition only exists at the thermal dynamical limit. Here, we propose an adiabatic scheme on a perturbed Ising spin model with the first order quantum phase transition. By employing the Landau-Zener anticrossing, we can not only encode the unknown parameter in the ground state but also tune the energy gap to control the evolution time of the adiabatic passage. We experimentally implement the adiabatic scheme on the nuclear magnetic resonance and show that the achieved precision attains the Heisenberg scaling. The advantages of the scheme---easy implementation, robust against the decay, tunable energy gap---are critical for practical applications of quantum metrology.
	\end{abstract}
	\maketitle
	
	\renewcommand{\thesubsection}{\arabic{subsection}}
	
	\par Quantum metrology, which makes use of the superposition and entanglement, can achieve far better precision than the classical schemes \cite{holevo2011,helstrom1969,cramer1999}. In the conventional scheme of quantum metrology, the estimation of an unknown parameter is typically achieved by first preparing a probe state, then letting the probe evolve under a dynamics that encodes the unknown parameter, the value of the parameter can then be estimated from the evolved state via a suitable measurement  \cite{holevo2011,helstrom1969,cramer1999,Giovannetti2011}.
	With an entangled probe state, quantum metrology can potentially enhance the precision from the classical shot noise limit, which scales as $N^{-1/2}$, to the Heisenberg limit, which scales as $N^{-1}$, here $N$ is the number of the probes \cite{holevo2011,helstrom1969,Giovannetti2011,GIOV06,GIOV04,Escher2011,Rafal2012,yuan2017quantum,HL1,HL4,HL5,HL6}. The classical shot noise limit and the Heisenberg limit can also be considered in terms of the evolution time, $T$, where the precision scales as $T^{-1/2}$ for the shot noise limit and $T^{-1}$ for the Heisenberg limit \cite{Rafal2014,Zhou2018}. For the conventional scheme, which consists of preparation, evolution and measurement, the ability to prepare highly entangled probe states or maintain a sufficiently long coherent evolution is essential to achieve a precision beyond the classical limit. This quantum advantage is not achievable in general for  systems subject to noise.
	
	Recently, the critical quantum metrology  \cite{rams2018,garbe2019,cri1,cri2,cri3,cri4,cri5,cri6,cri7,cri8,cri9,Ivanov2013,cri2021} has attracted increasing theoretical interest since it combines the advantages of the intrinsic robustness due to the adiabatic evolution \cite{robustAdi,rams2018} and high sensitivity near the critical point. Similar to the adiabatic quantum computation \cite{farhi2000,aharonov2008,AlbashRMP}, the adiabatic quantum metrology starts with the ground state of an initial Hamiltonian, which is easy to prepare, then evolves adiabatically to the ground state of the final Hamiltonian that encodes the unknown parameter. However, previous protocols typically consider systems with continuous quantum phase transitions, which only exists at the thermal dynamical limit, and the minimal energy gap at the critical point is also in general fixed which limits the speed of the adiabatic evolution. Such requirements impose great challenges on the experimental realization of the critical quantum metrology.



	In this Letter, we overcome these challenges by employing a perturbed two-spin system with a first-order phase transition where the energy gap can be tuned with the Landau-Zener anticrossing which controls the time required by the adiabatic passage. This can also be used to tune the tradeoff between the precision and the bandwidth of the estimation. We experimentally implement the scheme on a Nuclear Magnetic Resonance (NMR) quantum information processor and demonstrate a precision at the Heisenberg scaling. The adiabatic scheme is inherent robust against the decay since it remains at the ground state during the evolution, which we also verify with numerical simulations. As the first experiment demonstration of the the adiabatic quantum metrology, this opens the avenue for the exploration of the scheme on various physical systems for practical applications. 

	We implement the adiabatic scheme with an Ising model consisting of two spin-1/2, where the Hamiltonian is 
	\begin{eqnarray}
	\mathcal{H}_{\text{Ising}}=B_z(\sigma_{z}^1+\sigma_z^2)+\sigma_z^1\sigma_z^2,
	\end{eqnarray}
	here $\sigma_{z}^{i}$ is Pauli operator on the $i$-th spin, $B_z$ is the magnitude of the longitudinal magnetic field, which is the parameter to be estimated. The ground state of the Hamiltonian is given by
	\begin{eqnarray}
	|g(B_z)\rangle=\left\{\begin{array}{cc}
	|00\rangle;   &  B_z\leq -1,\\
	\frac{|01\rangle\pm|10\rangle}{\sqrt{2}};     & -1\leq B_z\leq 1, \\
	|11\rangle;   &  B_z\geq 1,
	\end{array}\right.
	\end{eqnarray}
	which has a sudden change at the critical points $B_z=\pm1$. 
	The ground state has a degeneracy of $2$ when $-1< B_z< 1$. The degeneracy, however, can be lifted by restricting to the symmetric triplet space \cite{peng2005}. Intuitively as the Hamiltonian is invariant under the exchange of the two spins, if the initial state is symmetric then the state will remain in the symmetric space during the evolution. We can thus only consider the symmetric states and the adiabatic evolution is only constrained by the energy gap of the effective Hamiltonian on the symmetric space \cite{Zhuang2020}. 
	
	The ground state on the symmetric space, however, still does not provide a precise information of $B_z$. To enable the estimation of $B_z$, we need a one-to-one correspondence between $B_z$ and the ground state. To achieve that we can add a small transverse field with $B_x\ll1$, 
	\begin{eqnarray}
	\label{ham}
	\widetilde{\mathcal{H}}_{\text{Ising}}=B_z(\sigma_{z}^1+\sigma_z^2)+B_x(\sigma_x^1+\sigma^2_x)+\sigma_z^1\sigma_z^2,
	\end{eqnarray}
	which preserves the symmetry. The transverse field transforms the singular jump at the critical point to a non-singular transition over a finite width. By tuning $B_x$, we can adjust the width and the rate of change near the critical point. This transverse field can also be used to tune the energy gap which determines the evolution time of the adiabatic passage.   

	We can implement the adiabatic evolution with an additional control field along the $z$-direction as 
	\begin{eqnarray}
	\label{ham}
	\widetilde{\mathcal{H}}_{}(t)=[B_z+B_c(t)](\sigma_{z}^1+\sigma_z^2)+B_x(\sigma_x^1+\sigma^2_x)+\sigma_z^1\sigma_z^2,
	\end{eqnarray}
	where $B_c(t)$ is the control field which adiabatically changes from a large value to zero. This preserves the symmetry of the evolution. In the experiment, $B_z$ and $B_c(t)$ is combined as a single field which is changed from a large value to some $B_z$, whose value is then estimated by proper measurements on the final state. As a proof of principle, we focus on the local estimation where $B_z$ is within a small neighborhood of a known value. 

 The precision of the local estimation can be characterized by the quantum Cramer-Rao bound(QCRB) \cite{holevo2011,cramer1999,helstrom1969,braunstein1994,braunstein1996} as
	\begin{equation}
	\delta B_z^2\geq \frac{1}{\nu F_Q},
	\end{equation}
	here $\nu$ is the number of repetitions of the experiment and $F_Q$ is the quantum Fisher information (QFI) \cite{holevo2011,cramer1999,helstrom1969} of the final state, $|\widetilde{g}\rangle$. The effective final Hamiltonian on the two lowest energy levels can be written as \cite{supp}
	\begin{equation}
	\mathcal{H}_{\text{eff}}=-|B_z|\mathbf{1}_2+(1-|B_z|)\sigma_z+\sqrt2B_x\sigma_x, 
	\end{equation}
where $\mathbf{1}_n$ denotes the $n\times n$ identity operator. When $B_z>0$, the ground state of the effective Hamiltonian can be written as
	\begin{eqnarray}
	\label{eq:ground}
	|\widetilde{g}(B_z)\rangle&=&-\text{sin}\frac{\theta}{2}|a\rangle+\text{cos}\frac{\theta}{2}|b\rangle,
	\end{eqnarray}
	where $|a\rangle=|11\rangle$, $|b\rangle=\frac{|01\rangle+|10\rangle}{\sqrt2}$ and  $\text{tan}\theta=\frac{\sqrt2B_x}{1-B_z}$ \cite{zhang2008}. 
The QFI of the ground state, which can be computed as
	\begin{eqnarray}
	\label{defQFI}
	F_{Q}(|\widetilde{g}\rangle)=4(\langle\partial_{B_{z}}\widetilde{g}|\partial_{B_{z}}\widetilde{g}\rangle-|\langle \widetilde{g}|\partial_{B_{z}}\widetilde{g}\rangle|^{2}),
	\end{eqnarray} can then be obtained as \begin{equation}
	\label{eq:QFI}
	F_{Q}(|\widetilde{g}\rangle)=\frac{2B_{x}^{2}}{[(1-B_{z})^{2}+2B_{x}^{2}]^{2}}.
	\end{equation}
Near the critical point, $F_Q(|\widetilde{g}\rangle)\approx \frac{1}{2B_x^2}$, which has a large value when $B_x\rightarrow 0$. To gauge the practical advantage near the critical point, however, we also need to evaluate the cost, which is the time, $T$, required for the adiabatic evolution. A QFI scales as $T^2$ corresponds to the Heisenberg scaling \cite{Rafal2014,Zhou2018,YuanTime} while a QFI scales as $T$ corresponds the shot noise limit.

    \begin{figure}[tb]
			\centering
    	\includegraphics[width=0.48\textwidth]{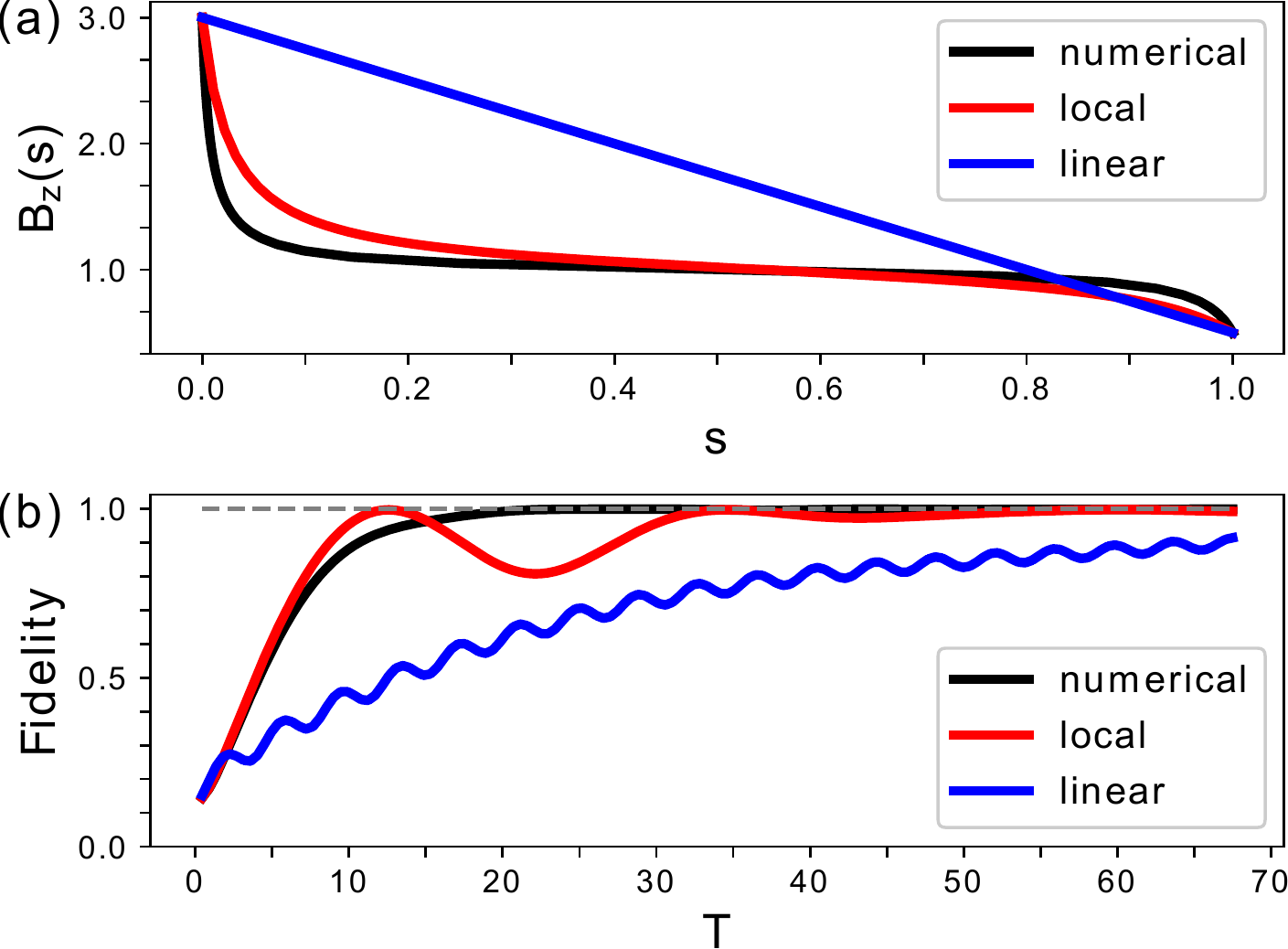}
		\caption{(a) The linear adiabatic path, the local adiabatic path and the numerically optimized path with $B_x=0.1$ and $B_z$ is adiabatically decreased from $3$ to $0.5$. (b) Fidelity between the adiabatically evolved state and the actual ground state under these three paths when the total evolution time varies, here the unit is $2/(\pi J)$.}\label{Fig1}
	\end{figure}

We consider the time required by the adiabatic evolution from an initial large $B_{z0}$ to the critical point, $B_{zc}\approx 1$. For the local precision limit where the field is within a small neighborhood of a known field, if the field to be estimated is not near the critical point, we can shift it by compensating it with an additional known field. For general unknown field that is not within a small neighborhood, this can be achieved through the two-step adaptive method \cite{FUJI06,Gill2005}. In this two-step method, the experiment is repeated where the first few experiments are used to obtain a rough estimation of the unknown field, with this rough estimation the field can then be shifted to near the critical point in the following experiments.

The adiabatic path can be described as  
\begin{equation}
\mathcal{H}_{\text{ad}}[A(s)]=[1-A(s)]\widetilde{\mathcal{H}}_{\text{Ising}}(B_{z0})+A(s)\widetilde{\mathcal{H}}_{\text{Ising}}(B_{zc}),    
\end{equation}
 where $\widetilde{\mathcal{H}}_{\text{Ising}}(B_{z0})$ is the initial Hamiltonian and  $\widetilde{\mathcal{H}}_{\text{Ising}}(B_{zc})$ is the final Hamiltonian, $s=t/T\in[0,1]$ is the normalized time, the function, $A(s)$, determines the adiabatic path with $A(0)=0$ and $A(1)=1$. 
	
The time required for the adiabatic path is determined by the adiabatic condition \cite{Jansen_2007}. The simplest adiabatic path is the linear path, which corresponds to $A(s)=s$. In this case the evolution time is of the order $\frac{1}{\Delta_{\min}^2}$ with $\Delta_{\min}$ as the minimal energy gap between the ground state and the first excited state \cite{farhi2000, supp}, 
In our case the energy gap is
\begin{equation}
	\Delta(s) = 2\sqrt{2B_x^2+[1-B_{z_0}+B_{z_0}A(s)-B_{zc}A(s)]^2},
\end{equation}
with $\Delta_{\min}=2\sqrt{2}B_x$. For the linear path we thus have $T\propto \frac{1}{B_x^2}$. The QFI, which is $F_Q(|\widetilde{g}\rangle)\approx \frac{1}{2B_x^2}$, then scales only linearly with $T$. More efficient adiabatic evolutions are required to go beyond the shot noise limit. One choice is the local adiabatic path, which adjusts the evolution speed according to the local energy gap as $\frac{dA(s)}{ds}=c\Delta^2(s)$ with $c$ as a constant \cite{local}. In this case the evolution time is of the order $\frac{1}{\Delta_{\min}}log\frac{1}{\Delta_{\min}}$ and the precision can go beyond the shot noise limit \cite{supp}. 

In our experiment we further optimize the adiabatic path numerically. The optimization is achieved as following: 1)first set a threshold on the fidelity, which is denoted as $P_c$(in our case $P_c=0.9999$); 2) start from $A(0)=0$, let $A_1$ be the minimal value such that $|\langle g(A_1)|e^{-i\mathcal{H}_{\text{ad}}(A_1)\tau}|g(0)\rangle|\leq P_c$, here $\tau$ is a fixed constant and $|g(A)\rangle$ is the ground state of $\mathcal{H}_{\text{ad}}(A)$; iteratively, we set $A_{i+1}$ as  the minimal value such that $|\langle g(A_{i+1})|e^{-i\mathcal{H}_{\text{ad}}(A_{i+1})\tau}|g(A_i)\rangle|\leq P_c$; 3) If $A_N\geq 1$, then set $A_N=1$ and the procedure terminates. An adiabatic path is then obtained with $A(\frac{i}{N})=A_i$. When $P_c$ is chosen sufficiently close to 1, the obtained path guarantees that the evolved state stays close to the ground state along the path and the time of this path is $\propto \frac{1}{B_x}$ \cite{supp}. This path is obtained from the fidelity directly, while the linear and the local paths are based on the energy gap which is related to the fidelity in an indirect way. We simulate the evolution of different adiabatic paths with the full Hamiltonian and it can be seen from Fig. \ref{Fig1} that the numerically obtained path shows a better performance. 
	
 \begin{figure}[tb]
		\begin{minipage}{1\linewidth}
			\centering
			\includegraphics[height=7cm,width=8cm]{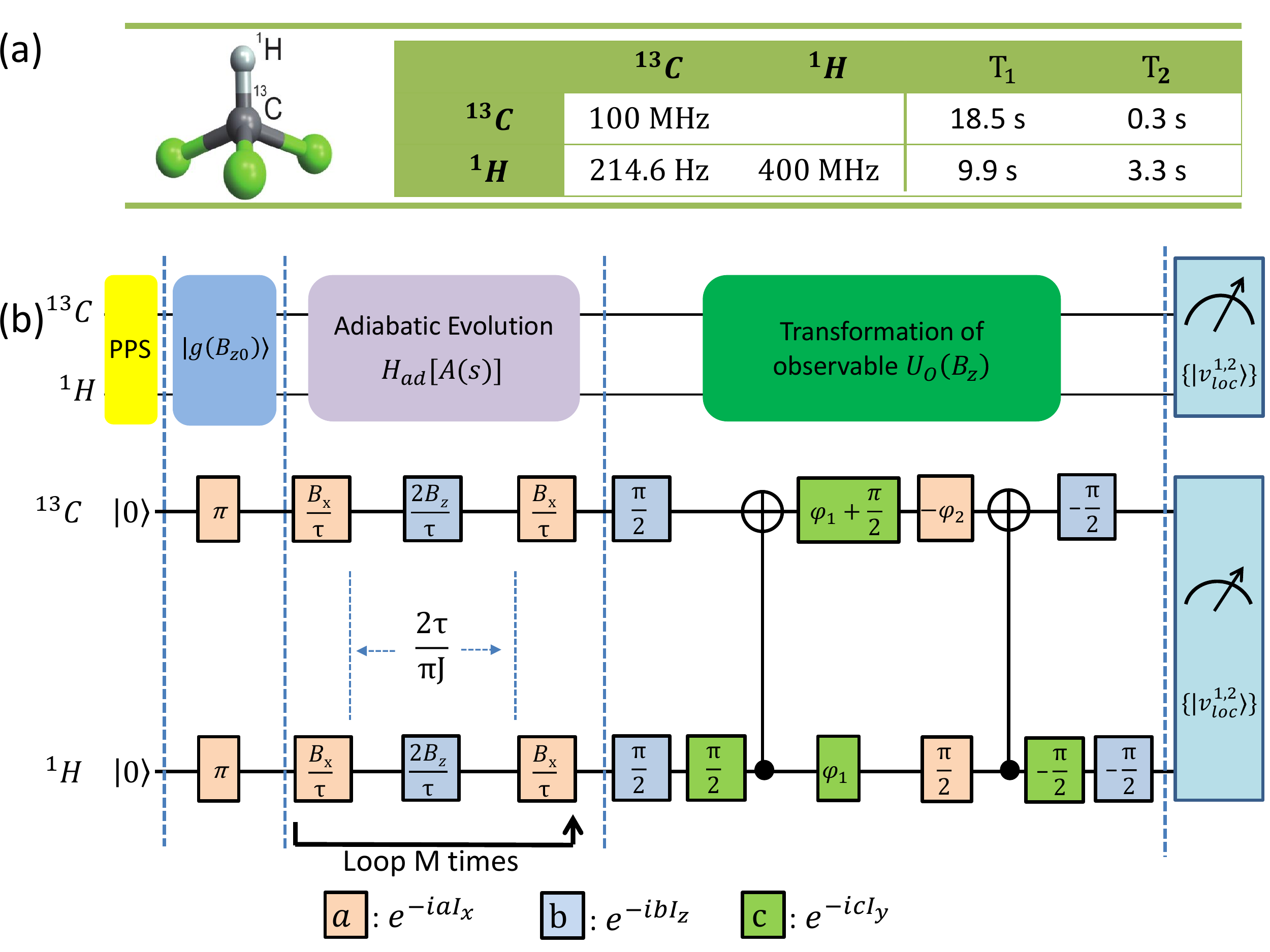}
		\end{minipage}
		\caption{(a) Molecular structure and relevant parameters of $^{13}$C labeled Chloroform. (b) Experimental scheme and quantum circuit for adiabatic quantum metrology on NMR. Here, $\varphi_1=\arctan [(B_z-1)/0]-\pi/4, \varphi_2=|\theta-\pi/2|$, where $\theta=\sqrt{2}B_x/(1-B_z)$. See supplementary material for the specific pulse sequence for implementing in experiment.}\label{Fig2}
	\end{figure}
 To saturate the QCRB, we need to perform the optimal measurement, which is the projective measurement on the eigenvectors of the symmetric logarithmic derivative (SLD). The SLD, denoted as $L$, can be obtained from the equation $\frac{\partial \rho(B_z)}{\partial B_z}=\frac{1}{2}[\rho(B_z)L+L\rho(B_z)]$  \cite{helstrom1969,braunstein1994,braunstein1996}.  
When $\rho(B_z)=|\widetilde{g}\rangle\langle \widetilde{g}|$, we have $L=2(|\partial_{B_z}\widetilde{g}\rangle\langle \widetilde{g}|+|\widetilde{g}\rangle\langle\partial_{B_z}\widetilde{g}|)$, whose eigenvectors are given by
	\begin{eqnarray}
	\label{Optv}
	|v_{\text{Opt}}^{1}\rangle&=&\sqrt{\frac{1-\text{sin}\theta}{2}}|11\rangle+\frac{\text{cos}\theta}{\sqrt{2(1-\text{sin}\theta)}}\frac{|01\rangle+|10\rangle}{\sqrt2}, \notag\\
	|v_{\text{Opt}}^{2}\rangle&=&-\sqrt{\frac{1+\text{sin}\theta}{2}}|11\rangle+\frac{\text{cos}\theta}{\sqrt{2(1+\text{sin}\theta)}}\frac{|01\rangle+|10\rangle}{\sqrt2},\notag
	\end{eqnarray}
	where $\theta$ takes the same value as in Eq. (\ref{eq:ground}). This optimal measurement depends on $B_z$, and in practice it can be implemented adaptively with the estimated value $B_z$ based on the previously accumulated measurement data \cite{ada1988,ada2005,FUJI06,Gill2005,HAYA08}.
	\begin{figure*}
        \includegraphics[width=1.0\textwidth]{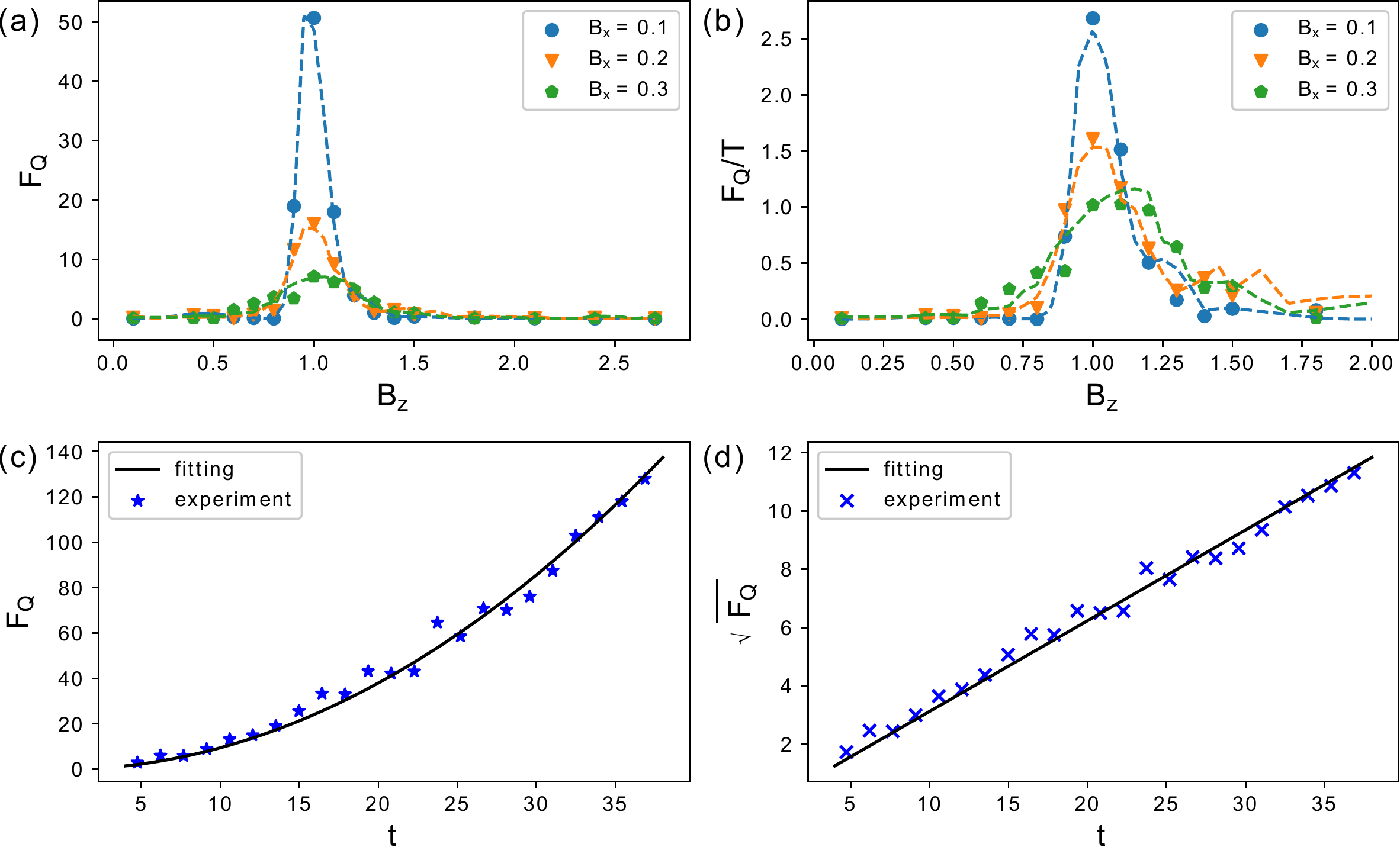}
		\caption{(a) Experimentally obtained QFI at different $B_z$ with $B_x$= 0.1, 0.2 and 0.3 (denoted by $\bullet, \triangledown, \pentagon$, respectively), along with the corresponding numerical simulations (denoted by dashed lines) together for comparation. (b) The obtained QFI per unit of time at different $B_z$ with $B_x$= 0.1, 0.2 and 0.3. (c) Experimentally obtained QFI (denoted by $\star$) and (d) its square root (denoted by $\times$) near the critical point with different adiabatic time $t$, which is achieved by tuning $B_x$. The solid lines in (c) and (d) represent the fittings with a quadratic function and  a linear function, respectively.}\label{Fig3}
	\end{figure*}

We implement the protocol on the Bruker Avance $\textrm{III}$ 400 MHz (9.4 T) spectrometer at the room temperature. The two nuclear spins, as shown in Fig. \ref{Fig2}, are $^{13}$C and $^1$H in the $^{13}$C-labeled chloroform which is dissolved in $d_6$ acetone. In the double-resonant rotating frame the natural Hamiltonian of this system is $ \frac {\pi}{2}J\sigma_z^1\sigma_z^2$, where $J=214.5$ Hz is the coupling strength. For convenience, we will take the time unit as $\frac{2}{\pi J}$ and write the Hamiltonian as 
$	\mathcal{H}_{\text{NMR}}=\sigma_z^1\sigma_z^2.$
The transverse field can be realized by the on-resonance radio-frequency pulse along the $x-$axis, and the vertical field can be generated with an appropriate offset of the transmitter’s frequency \cite{offset}.
    
    The initial state of the system is the pseudopure state (PPS), $\rho_{00} =\frac{1 - \epsilon}{4} \mathbf{1}_4+ \epsilon|00 \rangle \langle 00 |$   \cite{Peng_2001}, where $\epsilon\approx 10^{-5}$ represents the thermal polarization. We then prepare the ground state of the initial Hamiltonian  ${\mathcal{H}}_{\text{ad}}[A(0) = 0]=\widetilde{\mathcal{H}}_{\text{Ising}}(B_{z0})$ and adiabatically drive the system to $\mathcal{H}_{\text{ad}}[A(T) = 1] = \widetilde{\mathcal{H}}_{\text{Ising}}(B_{zf})$, where $B_{z0}$ is taken as $20$ in the experiment and $B_{zf}=0$.  
In the experiment, we use the trotterized adiabatic evolution with $M$ segments \cite{Sun_2020,Wu2002}, each with a duration $\Delta t=T/M$. 
For the numerical path $T= c/B_x$( in the experiment $c\approx 3.6$, see supplementary material for the details) and the step number is taken as $M = 100$. During each segment the field is approximated as a constant with $B_{z}[i]=[1-A(\frac{i}{M})]B_{z0}+A(\frac{i}{M})B_{zf}$ and the corresponding evolution, as shown in Fig. \ref{Fig2}(b), is generated via the trotterization as $U_{i}(\Delta t)=e^{-i\mathcal{H}_{\text{ad}}[A_i]\Delta t} = e^{-iB_x(\sigma_x^1+\sigma_x^2)\frac{\Delta t}{2}}e^{-i\{B_{z}[i](\sigma_z^1+\sigma_z^2)+\sigma_z^1\sigma_z^2\}\Delta t}e^{-iB_x(\sigma_x^1+\sigma_x^2)\frac{\Delta t}{2}} + O(\Delta t^3)$, where $e^{-iB_x(\sigma_x^1+\sigma_x^2)\frac{\Delta t}{2}}$ is realized by a strong resonant control pulse along the $x$-axis, $e^{-i[B_{z}[i](\sigma_z^1+\sigma_z^2)+\sigma_z^1\sigma_z^2]\Delta t}$ is realized  by a free evolution with an frequency offset $B_{z}[i]J/2$ \cite{peng2005}. 
        
In the experiment, we stop the adiabatic evolution at different $B_z$, which varies from $0.1$ to $2.7$, to get the ground state $|\widetilde{g}^{\text{exp}}(B_z)\rangle$, then perform the optimal projective measurements, $\{|v_{\text{Opt}}^{1}(B_{z})\rangle\langle v_{\text{Opt}}^{1}(B_{z})|, |v_{\text{Opt}}^{2}(B_{z})\rangle \langle v_{\text{Opt}}^{2}(B_{z})|\}$. In the experiment, only the local observables can be directly implemented. 
Specifically, the local observable implemented directly in our experiment is $\sigma_{x}^1\otimes \frac{1}{2} ({\bf{1}_2} - \sigma_{z}^2)$, whose eigenvectors are $|v_{\text{loc}}^1\rangle = \frac{1}{\sqrt{2}}( | 0 \rangle+| 1 \rangle)\otimes | 1 \rangle $ and $|v_{\text{loc}}^2\rangle = \frac{1}{\sqrt{2}}( | 0 \rangle - | 1 \rangle)\otimes | 1 \rangle$ with the corresponding eigenvalues $\lambda_1=1$ and $\lambda_2=-1$. To perform the optimal measurement, we first implement a unitary operation $U_O ({B}_{z})$ with $U_O ({B}_{z}) |v_{\text{Opt}}^m\rangle=|v_{\text{loc}}^m\rangle$, $m = 1,2$, then perform the local measurement. The detailed implementation of $U_O(B_{z})$ can be found in the supplemental material. In NMR the experimental signal corresponds to the average of the observable over an ensemble, which is given by $p_1(B_z)\lambda_1-p_2(B_z)\lambda_2$ with $p_m(B_z) = | \langle v_{\text{Opt}}^{m} \vert \widetilde{g}^{\text{exp}}(B_{z}) \rangle |^2 =  | \langle v_{\text{loc}}^{m} | U_O(B_{z})\vert  \widetilde{g}^{\text{exp}}(B_{z}) \rangle |^2 $. From the experimental signal, together with the condition $p_1(B_z)+p_2(B_z)=1$, we can get $p_1(B_z)$ and $p_2(B_z)$ respectively. 
   To get the Fisher information, $F_C^{opt}(B_z) = \frac{ [\partial_{B_z}p_1(B_z) ]^2}{p_1(B_z)} + \frac{ [\partial_{B_z}p_2(B_z) ]^2}{p_2(B_z)}$ (here $F_C$ is the classical Fisher information which equals to the QFI under the optimal measurement) \cite{helstrom1969}, we also need to get $\partial_{B_z}p_m(B_z)$ experimentally. This is achieved by the difference method, i.e., by repeating the experiment at two neighboring points, $B_z\pm\delta$, where $\delta$ is a small shift (taken as $0.03$ experimentally, see supplementary material for detail). The differentiation is then obtained as $\partial_{B_z}p_m(B_z) \approx \frac{p_m(B_z+\delta)-p_m(B_z-\delta)}{2\delta}$. 

       \par The experiment is repeated under $B_x=$ $0.1,0.2$, $0.3$, where for each $B_x$, $B_z$ is varied non-uniformly from 0.1 to 2.7. As shown in Fig. \ref{Fig3}(a), under all $B_x$, the QFI around the critical point is significantly higher than the QFI away from the critical point. The total relative deviation of the experimental data from the numerical simulations is about 8.8\% (see supplementary material). To show the practical advantage, we also plot the QFI per unit of time, $F_Q(T)/T$, in Fig. \ref{Fig3}(b), which is also significantly higher around the critical point. This shows the critical point indeed provides an advantage in quantum metrology.        
       
	\par To demonstrate the scaling of the QFI with respect to the time, we perform another set of experiments where we adiabatically evolve the system from $B_{z0}=20$ to the critical point, $B_{zc}=1$, with $B_x$ tuned at different values to control the evolution time (since $T \propto 1/B_x$ for the numerical adiabatic path). By experimentally obtaining the QFI under different $B_x$, we plot the relation of the QFI with the evolution time. As it can be seen from Fig. \ref{Fig3}(c), the QFI scales quadratically with the time. The total relative deviation of the experimental result from the numerical simulation is about 5.1\% (see supplementary material). To better illustrate the scaling, we also plot $\sqrt{F_Q}$ with respect of the time in Fig. \ref{Fig3}(d), where $\sqrt{F_Q}\propto T$ can be clearly seen. The coefficient of determination \cite{R2} of the linear fitting is 98.6\%, and the slope of the fitted line is 0.31 with an uncertainty of 0.0032. This shows clearly that the adiabatic scheme achieves the Heisenberg scaling near the critical point. 
	The detailed analysis of its performance, including bandwidth and accuracy, is provided in the supplemental material \cite{supp}. We also numerically compare our protocol with the standard scheme of quantum metrology at the presence of the decay \cite{supp}, and show it can surpass the standard scheme due to its robustness against decay. 
	
	\par In summary, we performed the first experimental adiabatic quantum metrology on the NMR quantum processor and demonstrated that with the optimized adiabatic path the precision can achieve the Heisenberg scaling. This scheme is easier to implement and robust against the decays, which opens a promising experimental path for practical quantum metrology. We expect this will lead to the investigation of the adiabatic quantum metrology for many practical applications on various physical systems, such as NV center \cite{NVc} and cold atoms \cite{coldatoms}.
The adiabatic quantum metrology also connects the precision limit to the speed of the adiabatic evolution, various bounds in quantum metrology thus can also be used to study the speed limit of the adiabatic passage under noisy evolutions, which is another interesting direction to pursue. 
\begin{acknowledgements}
This work is supported by National Key Research and Development Program of China (Grant No. 2018YFA0306600), the National Natural Science Foundation of China (Grant No. 11661161018, Grant No. 11927811), Anhui Initiative in Quantum Information Technologies (Grant No. AHY050000), Research Grants Council of Hong Kong (GRF No. 14308019) and the Research Strategic Funding Scheme of The Chinese University of Hong Kong (No. 3133234).
\end{acknowledgements}
	\bibliographystyle{apsrev4-2}
	\bibliography{reference}
\end{document}


\title{Experimental Adiabatic Quantum Metrology with the Heisenberg scaling: Supplemental Material}

	\date{\today}
\author{Ran Liu}
\thanks{These two authors contributed equally}
\affiliation{Hefei National Laboratory for Physical Sciences at the Microscale and Department of Modern Physics, University of Science and Technology of China, Hefei 230026, China}
\affiliation{CAS Key Laboratory of Microscale Magnetic Resonance, University of Science and Technology of China, Hefei 230026, China}
\affiliation{Synergetic Innovation Center of Quantum Information and Quantum Physics, University of Science and Technology of China, Hefei 230026, China}
\author{Yu Chen}
\thanks{These two authors contributed equally}
\affiliation{Department of Mechanical and Automation Engineering, The Chinese University of Hong Kong, Shatin, Hong Kong SAR, China}
	\author{Min Jiang}
\affiliation{Hefei National Laboratory for Physical Sciences at the Microscale and Department of Modern Physics, University of Science and Technology of China, Hefei 230026, China}
\affiliation{CAS Key Laboratory of Microscale Magnetic Resonance, University of Science and Technology of China, Hefei 230026, China}
\affiliation{Synergetic Innovation Center of Quantum Information and Quantum Physics, University of Science and Technology of China, Hefei 230026, China}
\author{Xiaodong Yang}
\affiliation{Hefei National Laboratory for Physical Sciences at the Microscale and Department of Modern Physics, University of Science and Technology of China, Hefei 230026, China}
\affiliation{CAS Key Laboratory of Microscale Magnetic Resonance, University of Science and Technology of China, Hefei 230026, China}
\affiliation{Synergetic Innovation Center of Quantum Information and Quantum Physics, University of Science and Technology of China, Hefei 230026, China}
	\author{Ze Wu}
\affiliation{Hefei National Laboratory for Physical Sciences at the Microscale and Department of Modern Physics, University of Science and Technology of China, Hefei 230026, China}
\affiliation{CAS Key Laboratory of Microscale Magnetic Resonance, University of Science and Technology of China, Hefei 230026, China}
\affiliation{Synergetic Innovation Center of Quantum Information and Quantum Physics, University of Science and Technology of China, Hefei 230026, China}
\author{Yuchen Li}
\affiliation{Hefei National Laboratory for Physical Sciences at the Microscale and Department of Modern Physics, University of Science and Technology of China, Hefei 230026, China}
\affiliation{CAS Key Laboratory of Microscale Magnetic Resonance, University of Science and Technology of China, Hefei 230026, China}
\affiliation{Synergetic Innovation Center of Quantum Information and Quantum Physics, University of Science and Technology of China, Hefei 230026, China}
\author{Haidong Yuan}
\email{hdyuan@mae.cuhk.edu.hk}
\affiliation{Department of Mechanical and Automation Engineering, The Chinese University of Hong Kong, Shatin, Hong Kong SAR, China}
\author{Xinhua Peng}
\email{xhpeng@ustc.edu.cn}
\affiliation{Hefei National Laboratory for Physical Sciences at the Microscale and Department of Modern Physics, University of Science and Technology of China, Hefei 230026, China}
\affiliation{CAS Key Laboratory of Microscale Magnetic Resonance, University of Science and Technology of China, Hefei 230026, China}
\affiliation{Synergetic Innovation Center of Quantum Information and Quantum Physics, University of Science and Technology of China, Hefei 230026, China}
\author{Jiangfeng Du}
\affiliation{Hefei National Laboratory for Physical Sciences at the Microscale and Department of Modern Physics, University of Science and Technology of China, Hefei 230026, China}
\affiliation{CAS Key Laboratory of Microscale Magnetic Resonance, University of Science and Technology of China, Hefei 230026, China}
\affiliation{Synergetic Innovation Center of Quantum Information and Quantum Physics, University of Science and Technology of China, Hefei 230026, China}
	\maketitle
		\section{Effective Hamiltonian of Ising Model}
	The ground state of $\mathcal{H}_{\text{Ising}}=B_z(\sigma_z^1+\sigma_z^2)+\sigma_z^1\sigma_z^2$ in the symmetric space is given by
	\begin{eqnarray}
	|g(B_z)\rangle=\left\{\begin{array}{cc}
	|00\rangle;   &  B_z\leq -1,\\
	\frac{|01\rangle\pm|10\rangle}{\sqrt{2}};     & -1\leq B_z\leq 1, \\
	|11\rangle;   &  B_z\geq 1,
	\end{array}\right.
	\end{eqnarray}
where the corresponding eigen-energies are given by $1+2B_z$, $-1$ and $1-2B_z$ respectively.
After adding a small transverse field, the Hamiltonian changes to $\widetilde{\mathcal{H}}_{\text{Ising}}=B_z(\sigma_z^1+\sigma_z^2)+\sigma_z^1\sigma_z^2+B_x(\sigma_x^1+\sigma_x^2)$. When $B_x \ll 1$, the ground states are very close to those of $\mathcal{H}_{\text{Ising}}$, except in the vicinity of the critical points, where the transverse field mixes them, thus avoiding the level crossing. In this region, it is sufficient to consider the two lowest energy states $\{|a\rangle:=|11\rangle, |b\rangle:=\frac{|01\rangle+|10\rangle}{\sqrt{2}}\}$. They form a two-level system that can be described by the effective Hamiltonian
\begin{equation}
	\mathcal{H}_{\text{eff}}(B_z)  = -|B_z|\mathbf{1}_2 + (1-|B_z|)\sigma_z + \sqrt{2}B_x \sigma_x,
	\label{Heff}
	\end{equation}
where $\mathbf{1}_2$ denotes the $2\times2$ unit operator. We can then have the corresponding ground state for $B_z>0$ as
\begin{eqnarray}
	|\widetilde{g}(B_z)\rangle&=&-\text{sin}\frac{\theta}{2}|a\rangle+\text{cos}\frac{\theta}{2}|b\rangle,
	\label{gBz}
	\end{eqnarray}
	where $\text{tan}\theta=\frac{\sqrt2B_x}{1-B_z}$ with $\theta\in[0,\pi]$.
	
	\section{Evolution time of the local adiabatic passage}\label{appscal}
The adiabatic evolution is realized as
	\begin{equation}\label{Hadi}
	\mathcal H_{\text{ad}}[A(s)]=[1-A(s)]\widetilde{\mathcal H}_{\text{Ising}}(B_{z0})+A(s)\widetilde{\mathcal H}_{\text{Ising}}(B_{zf}), 
	\end{equation}
with the initial Hamiltonian $\widetilde{\mathcal H}_{\text{Ising}}(B_{z0})= B_{z_0}(\sigma_z^1+\sigma_z^2)+B_x(\sigma_x^1+\sigma_x^2)+\sigma_z^1\sigma_z^2$ and the final Hamiltonian $\widetilde{\mathcal H}_{\text{Ising}}(B_{zf})=B_{zf}(\sigma_z^1+\sigma_z^2)+B_x(\sigma_x^1+\sigma_x^2)+\sigma_z^1\sigma_z^2$. When $t$ goes from $0$ to $T$, $A(s)$ changes from $A(0)=0$ to $A(1)=1$. Here $s=\frac{t}{T}\in [0,1]$.  
The evolution time $T$ of the adiabatic path typically depends on the energy gap between the ground state and the first excited state. When $B_x \ll 1$, the adiabatic evolution Hamiltonian can also be written as an effective Hamiltonian on the two lowest energy eigenstates: 

\begin{equation}
	\mathcal H_{\mathrm{\text{ad}}}^{\text{eff}}[A(s)] = [1-A(s)]\mathcal{H}_{\text{eff}}(B_{z0})+A(s)\mathcal{H}_{\text{eff}}(B_{zf}),
\end{equation}
 and the energy gap can be obtained as
\begin{equation}\label{gap}
	\Delta[A(s)] = 2\sqrt{2B_x^2+[1-B_{z0}+B_{z0}A(s)-B_{zf}A(s)]^2}.
\end{equation}
At the critical point $B_{zc}=1$, the energy gap achieves the minimum $\Delta_{\min}=2\sqrt{2}B_x$. 
	
The time required for the adiabatic path is given by
the adiabatic condition \cite{Jansen_2007}
	\begin{equation}
	\label{eq:linear}
	T \gg 2\max_s \frac{\|\partial_s \mathcal H_{\mathrm{\text{ad}}}^{\prime}\|}{\Delta^2} + \int_0^1 \left(\frac{\|\partial_s^2 \mathcal H_{\mathrm{\text{ad}}}^{\prime}\|}{\Delta^2}+\frac{\|\partial_s \mathcal H_{\mathrm{\text{ad}}}^{\prime}\|^2}{\Delta^3}\right)ds.
	\end{equation}
Here, $\|...\|$ is Frobenius norm. For an $m\times n$ matrix $H$, its Frobenius norm is defined as $\|H\|\equiv\sqrt{\sum_{i=1}^m\sum_{j=1}^n|H_{ij}|^2}$.
For the linear adiabatic path, i.e., $A(s) = s$, we have $\|\partial_s \mathcal H_{\mathrm{\text{ad}}}^{\prime}\| = 2|B_{z0}-1|$, and Eq. (\ref{eq:linear}) goes as
	\begin{equation}
	T \gg \frac{2|B_{z0}-1|}{\Delta_{\min}^2} \sim \frac{1}{B_x^2}.
	\end{equation}
Therefore, the QFI 

$$
F_Q=\frac{2B_x^2}{[(1-B_z)^2+2B_x^2]^2} \sim \frac{1}{B_x^2} \propto T, 
$$
which only achieves the shot noise limit.
	
The local adiabatic passage is given by \cite{Tscal2}
	\begin{equation}
	\partial_s A(s) = c \Delta^2[A(s)],
	\end{equation}
	with $c =\int_s\Delta^{-2}(s)\partial_s A(s)ds=\int_0^1 \Delta^{-2}(A(s))dA(s)$ as the normalization factor that ensures $A(1)=1$.
	In this case we have
	\begin{equation}
	\aligned
	& \|\partial_s \mathcal H_{\mathrm{\text{ad}}}^{\prime}\| = 2(B_{z_0}-1)\|\partial_s A(s)\| = 2(B_{z_0}-1)c\Delta^2, \\
	& \|\partial_s^2 \mathcal H_{\mathrm{\text{ad}}}^{\prime}\| = |4(B_{z_0}-1)c\Delta \partial_s \Delta|.
	\endaligned
	\end{equation}
	Thus
	\begin{equation}
	\aligned
	& \int_0^1 \left(\frac{\|\partial_s^2\mathcal H_{\mathrm{\text{ad}}}^{\prime}\|}{\Delta^2}+\frac{\|\partial_s \mathcal H_{\mathrm{\text{ad}}}^{\prime}\|^2}{\Delta^3}\right)ds \\
	=& \int_0^1 -\frac{4(B_{z0}-1)c\Delta\partial_s\Delta}{\Delta^2} + \frac{4(B_{z_0}-1)^2c^2\Delta^4}{\Delta^3} ds\\
	= & 4(B_{z_0}-1)c\log\frac{\Delta(0)}{\Delta(1)} + \frac{4(B_{z0}-1)^2c\Delta^2\partial_sA(s)}{\Delta^3} ds\\
	= & 4(B_{z_0}-1)c\log\frac{\Delta(0)}{\Delta(1)} + 4(B_{z_0}-1)^2c\int_0^1 \Delta^{-1}(A(s))dA(s).
	\endaligned
	\end{equation}
Then
	\begin{equation}
	T_{\mathrm{local}} \gg 4(B_{z_0}-1)c + 4(B_{z_0}-1)c\log \frac{\Delta(0)}{\Delta(1)} + 4(B_{z_0}-1)^2c\int_0^1 \Delta^{-1}(A(s))dA(s).
	\end{equation}
Since
	\begin{equation}
	\aligned
	& c = \int_{0}^1 \Delta^{-2}(A(s))dA(s) = \frac{1}{4}\int_0^1 \frac{dA(s)}{2B_x^2+(B_{z_0}-1)^2(A(s)-1)^2} = \frac{1}{4\sqrt{2}}\frac{\mathrm{arccot}\frac{\sqrt{2}B_x}{B_{z_0}-1}}{(B_{z_0}-1)B_x}, \\
	& \int_{0}^1 \Delta^{-1}(A(s))dA(s) = -\frac{\log \left(\frac{\sqrt{2 B_x^2+(B_{z_0}-1)^2}-B_{z_0}+1}{\sqrt{2 B_x^2+(B_{z_0}-1)^2}+B_{z_0}-1}\right)}{4 (B_{z_0}-1)} \sim \log \frac{1}{B_x},
	\endaligned
	\end{equation}
and we note that $\frac{\mathrm{arccot} x}{x} \sim \frac{\pi/2}{x}$, which is in the order of $\frac{1}{x}$, we have $c\sim\frac{1}{B_x}$. Thus
	\begin{equation}
	T_{\mathrm{local}} \gg \frac{1}{B_x} + \frac{C}{B_x}\log \frac{1}{B_x} \sim \frac{1}{Bx}\log\frac{1}{Bx}.
	\end{equation}
As $F_Q\sim \frac{1}{B_x^2}$, it is easy to see $\lim_{B_x\rightarrow 0}\frac{F_Q}{T^{2-\epsilon}_{local}}\rightarrow \infty$ for any $\epsilon>0$, i.e., $F_Q$ is close to th Heisenberg scaling under the local adiabatic passage.

	\section{Numerical Optimization of adiabatic paths with iterative algorithm}\label{appopt}
	\par For the adiabatic path, $\mathcal{H}_{ad}[A(s)]=[1-A(s)]\mathcal{H}_i+A(s)\mathcal{H}_f$, the numerical path is obtained through the following steps:
	\begin{itemize}
		\setlength{\itemsep}{0pt}
		\setlength{\parsep}{0pt}
		\setlength{\parskip}{0pt}
		\item 1: Set the step size for the change of $A(s)$ as $\Delta A\approx 0.001$, the step size of the evolution time as $\Delta t$, and a threshold for the fidelity as $P_c$, which is taken as $0.9999$ in our case. 
		\item 2: Start from the ground state of $\mathcal{H}_i$, $|g(0)\rangle$, increase $A(s)$ by $\Delta A$ and evolve the state under the Hamiltonian for $\Delta t$ units of time, which leads to a state $|\phi_0(\Delta A)\rangle=e^{-i\mathcal{H}_{ad}(\Delta A)\Delta t}|g(0)\rangle$. Compute the fidelity between the state and the instantaneous ground-state $|g(\Delta A)\rangle$, which is $P_t=|\langle g(\Delta A)|\phi_0(\Delta A)\rangle|$. If $P_t\ge P_c$, continue to increase $A(s)$ by $\Delta A$ until $P_t=|\langle g[(n_1+1)\Delta A]|e^{-i\mathcal{H}_{ad}[(n_1+1)\Delta A]\Delta t}|g(0)\rangle|< P_c$. Set $A_1=n_1\Delta A$.
		\item 3: Similarly, start from the ground state of $\mathcal{H}_{ad}(A_1)$, and increase $A(s)$ till $P_t=|\langle g[A_1+(n_2+1)\Delta A]|e^{-i\mathcal{H}_{ad}[A_1+(n_2+1)\Delta A]\Delta t}|g(A_1)\rangle|< P_c$. Set $A_2=A_1+n_2\Delta A=n_1\Delta A+n_2\Delta A$. Similarly, we can get $A_3, A_4,\cdots ,A_n,\cdots$.
		\item 4: When $A_N=\sum_{i=1,2,...,N}n_i\Delta A\ge1$, set $A_N=1$.
	\end{itemize}
The adiabatic path is then obtained with $A(\frac{i}{N})=A_i$. We provide an analysis on the time scale of this path. Note that $A_{i+1}$ is determined from  $P_t=|\langle g(A_{i+1})|e^{-i\mathcal{H}_{ad}[A_{i+1}\Delta t]}|g(A_i)\rangle|$ with $A_{i+1}=A_i+n_{i+1}\Delta A$ such that $P_t\approx P_c$. Since $P_c$ is very close to $1$, the difference between $A_{i+1}$ and $A_i$ is very small. We can thus use the perturbation theory to compute the evolution of the ground state of $\mathcal H_{ad}(A_i)$ under the Hamiltonian $\mathcal{H}_{ad}(A_{i+1}) = \mathcal{H}_{ad}(A_i) + n_{i+1}\Delta A (\mathcal{H}_f - \mathcal{H}_i)$ and obtain 
	\begin{equation}
	P_t = 1 - \frac{n_{i+1}^2\Delta A^2}{2} \sum_{k \neq g} \frac{\langle k|\mathcal{H}_f - \mathcal{H}_i|g\rangle \langle g | \mathcal{H}_f - \mathcal{H}_i|k\rangle }{(E_k - E_g)^2} := 1 - \frac{n_{i+1}^2\Delta A^2}{f(A_i)}
	\end{equation}
	where $|g\rangle$ denotes the ground state and $\{|k\rangle\}$ denote the excited states of $\mathcal H_{ad}(A_i)$. As we can see from $f(A_i)$, where 
	\begin{equation}
		f(A_i)=\frac{2}{\sum_{k \neq g} \frac{\langle k|\mathcal{H}_f - \mathcal{H}_i|g\rangle \langle g | \mathcal{H}_f - \mathcal{H}_i|k\rangle }{(E_k - E_g)^2}},
	\end{equation}
	it not only includes the ground state and the first excited state, but also the higher excited states. This is one of the reasons why this numerical optimized path has better performance than the local adiabatic passage as the local adiabatic passage only makes use of the ground state and the first excited state. The maximal $n_{i+1}\Delta A$ is restricted by $P_t \approx P_c$, which leads to $n_{i+1}\Delta A\approx \sqrt{(1-P_c)f(A_i)}$.
	According to the adiabatic theorem,
\begin{equation}
    T \gg \max_{s\in [0,1]} \frac{|\langle k|\partial_s \mathcal H_{ad}[A(s)]|g\rangle|}{(E_k-E_g)^2}, \forall k \ge 1
\end{equation}
here
\begin{equation}
    \aligned 
    \partial_s \mathcal H_{ad}[A(s)]|_{s=\frac{i}{N}} \approx & \frac{\mathcal H_{ad}(A_{i+1})-\mathcal H_{ad}(A_i)}{\Delta s} \\ 
    = & \frac{n_{i+1}\Delta A (\mathcal{H}_f - \mathcal{H}_i)}{\Delta s}\\ 
    = & \frac{\sqrt{(1-P_c)f(A_i)}(\mathcal{H}_f-\mathcal{H}_i)}{\Delta s},
    \endaligned
\end{equation}
we thus have
\begin{equation}
\aligned 
    \frac{|\langle k|\partial_s \mathcal H_{ad}[A(s)]|g\rangle|}{(E_k-E_g)^2} \approx & \frac{\sqrt{1-P_c}|\langle k |\mathcal{H}_f-\mathcal{H}_i|g\rangle |}{\Delta s}\frac{\sqrt{f}}{(E_k-E_g)^2} \\  \le & \frac{\sqrt{2(1-P_c)}|\langle k |\mathcal{H}_f-\mathcal{H}_i|g\rangle |}{\Delta s} \frac{1}{|\langle k|\mathcal{H}_f-\mathcal{H}_i|g\rangle |(E_k-E_g)} \\ 
    =&\frac{\sqrt{2(1-P_c)}}{\Delta s} \frac{1}{(E_k-E_g)}\\
   \le &\frac{\sqrt{2(1-P_c)}}{\Delta s} \frac{1}{(E_1-E_g)}.
\endaligned
\end{equation}
Here, $E_1$ is the eigenvalue of the first excited state of $\mathcal H_{ad}(A_i)$. The time required for the numerical optimized path is then in the order of $\frac{1}{E_1-E_g}$ (i.e., $T\sim\frac{c}{B_x}$), which is consistent with the numerical simulation in the main text.

\section{Experimental adiabatic paths}

It is difficult to experimentally realize so many segments in the numerical adiabatic path optimized above, consequently, we use the linear interpolation to construct $M+1$ segments for the experimental implementation, as shown in Fig. \ref{supFig6}(b), (d), (f)(correspond to $B_x=0.1,0.2,0.3$, respectively). Here the adiabatic evolution was realized with
$M +1$ discrete steps, in the $i$-th step the evolution is governed by the Hamiltonian $\mathcal H_{\text{ad}}(A[i])$ with $A[i] = A(\frac{i}{M})$, which corresponds to a constant field, $B_{z}[i]=[1-A(\frac{i}{M})]B_{z0}+A(\frac{i}{M})B_{zf}$.

The evolution of each segment $U_i(\Delta t)=e^{-i[B_{z}[i](\sigma_{z}^1+\sigma_z^2)+B_x(\sigma_x^1+\sigma^2_x)+\sigma_z^1\sigma_z^2]\Delta t}$, here $\Delta t=T/(M+1)$, is implemented approximately as $U_i^{exp}(\Delta t)=e^{-iB_x(\sigma_x^1+\sigma_x^2)\Delta t/2}e^{-i[B_{z}[i](\sigma_z^1+\sigma_z^2)+\sigma_z^1\sigma_z^2]\Delta t}e^{-iB_x(\sigma_x^1+\sigma_x^2)\Delta t/2}$. 
Here $\Delta t$ and $M$ need to be optimized to satisfy (i)  $\Delta t$ is sufficient small so that $U_i(\Delta t)$ and $U_i^{exp}(\Delta t)$ is sufficiently close for all $B_{z}[i]$; (ii) $\Delta t$ is not too small so that the number of the total segments, $M+1$, is not too big; (iii) the total time $(M+1)\Delta t$ is not too large at the presence of the decoherence. We include the relaxation effect in the optimization, where for $^{13}$C we take $T_1$ and $T_2$ as $18.5$s and $0.2$s respectively, and for $^1$H we take $T_1$ and $T_2$ as $9.9$s and $0.6$s respectively. 
\begin{figure}[htb]
	\begin{center}
		\includegraphics[height=11cm,width=1\linewidth]{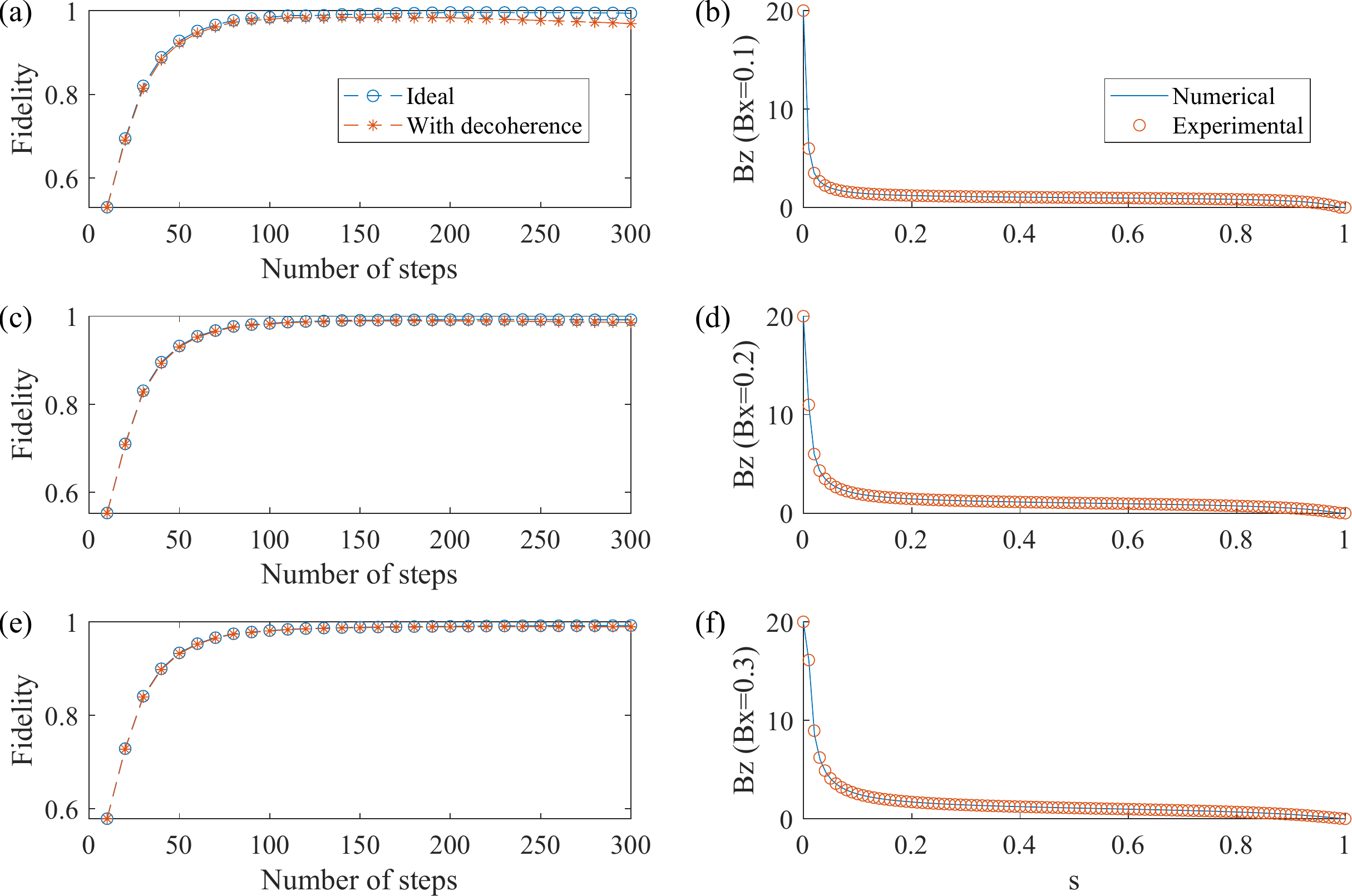} %
		
		\caption{Numerical simulations of the average fidelity during the adiabatic passage versus the number of steps $M+1$ for (a) $B_x=0.1$,  (c) $B_x=0.2$ and  (e) $B_x=0.3$. (b), (d), (f), the corresponding experimental adiabatic paths $B_z[i]$ with $i=0,1,M=99$ (denoted by the red circles), along with the numerical adiabatic paths optimized by iterative algorithm (denoted by the blue solid lines).}\label{supFig6}
	\end{center}
\end{figure}

In Fig.\ref{supFig6}(a) we plot the average overlap between the generated states and the ideal ground states for the case of $B_x=0.1$, 
where the overlap is define as 
\begin{eqnarray}
	F(\rho_1,\rho_2)=\frac{Tr(\rho_1\rho_2)}{\sqrt{Tr(\rho_1^2)Tr(\rho_2^2)}}.
\end{eqnarray}
Upon the optimization $\Delta t$ is chosen as $0.36$ (in the unit of $2/\pi J$) for which the fidelity between $U_i(\Delta t)$ and $U_i^{exp}(\Delta t)$ is above $99.8\%$ for all $B_{z}[i]$, $M+1$ is chosen as $100$ which achieves the maximal average fidelity, as shown in Fig. \ref{supFig6}(a). The total adiabatic evolution time is then $T=M\Delta t=36$. Since for the numerical path we have $T\approx \frac{
c}{B_x}$, then $c\approx T\times B_x=3.6$ in this case.

Similarly, for $B_x=0.2,0.3$, we take the same constant $c=3.6$ and the same number of segments(i.e, $M +1 = 100$) for consistency. The total adiabatic evolution time for $B_x=0.2$ is then $T = 3.6/0.2=18$ and for $B_x=0.3$, $T = 3.6/0.3=12$. The numerical simulations and the corresponding experimental adiabatic paths are shown in Fig. \ref{supFig6} (c) - (f).  


	\section{Realization of Optimal measurement}\label{apptrans}
	In the experiment, the direct local observable is 
	\begin{eqnarray}
	O_{loc}=\sigma_x\otimes|1\rangle\langle1|,
	\end{eqnarray}
which can be diagonalized as
	\begin{eqnarray}
	O_{loc}=\sum_{1\le m\le4}\lambda_m|v_{loc}^m\rangle\langle v_{loc}^m|,
	\end{eqnarray}
	here $\lambda_1=1,\lambda_2=-1,\lambda_3=\lambda_4=0,|v_{loc}^1\rangle=|+,1\rangle,|v_{loc}^2\rangle=|-,1\rangle,|v_{loc}^3\rangle=|+,0\rangle,|v_{loc}^4\rangle=|-,0\rangle$ with $|\pm\rangle=\frac{|0\rangle\pm|1\rangle}{\sqrt 2}$.
	\par The unitary operator that transforms the optimal observable to the local observable can be constructed as
	\begin{eqnarray}
	U_O(B_z)
	=\sum_{m=1}^4|v_{loc}^m \rangle\langle v_{Opt}^m(B_z)|
	\end{eqnarray}
	where $|v_{Opt}^1(B_z)\rangle$ and $|v_{Opt}^2(B_z)\rangle$ are the basis of the optimal measurement given in the main text, and $|v_{Opt}^{3}(B_z)\rangle$ and $|v_{Opt}^{4}(B_z)\rangle$ are two additional vectors to form a complete orthonormal basis. Hence, the effective optimal observable we employed can be expressed as
	\begin{eqnarray}\label{OptM}
		O_{Opt}(B_z)=\sum_{1\le m\le4}\lambda_m|v_{Opt}^m(B_z)\rangle\langle v_{Opt}^m(B_z)|.
	\end{eqnarray}
	Without loss of generality, we can take
	
	\begin{eqnarray}
	\label{Optv}
	|v_{Opt}^{1}({B}_z)\rangle&=&\sqrt{\frac{1-\text{sin}\theta}{2}}|11\rangle+\frac{\text{cos}\theta}{\sqrt{2(1-\text{sin}\theta)}}\frac{|01\rangle+|10\rangle}{\sqrt2}, \notag\\
	|v_{Opt}^{2}({B}_z)\rangle&=&-\sqrt{\frac{1+\text{sin}\theta}{2}}|11\rangle+\frac{\text{cos}\theta}{\sqrt{2(1+\text{sin}\theta)}}\frac{|01\rangle+|10\rangle}{\sqrt2}, \notag\\
	|v_{Opt}^{3}({B}_z)\rangle&=&\sqrt{\frac{1-\text{sin}\theta}{2}}|00\rangle+\frac{\text{cos}\theta}{\sqrt{2(1-\text{sin}\theta)}}\frac{|01\rangle-|10\rangle}{\sqrt2}, \notag\\
	|v_{Opt}^{4}({B}_z)\rangle&=&-\sqrt{\frac{1+\text{sin}\theta}{2}}|00\rangle+\frac{\text{cos}\theta}{\sqrt{2(1+\text{sin}\theta)}}\frac{|01\rangle-|10\rangle}{\sqrt2} \notag\\
	\end{eqnarray}	
with $\theta=\sqrt{2}B_x/(1-B_z)$.
	\par 
		\begin{figure}[htb]
		\begin{center}
			\includegraphics[height=12cm,width=0.9\linewidth]{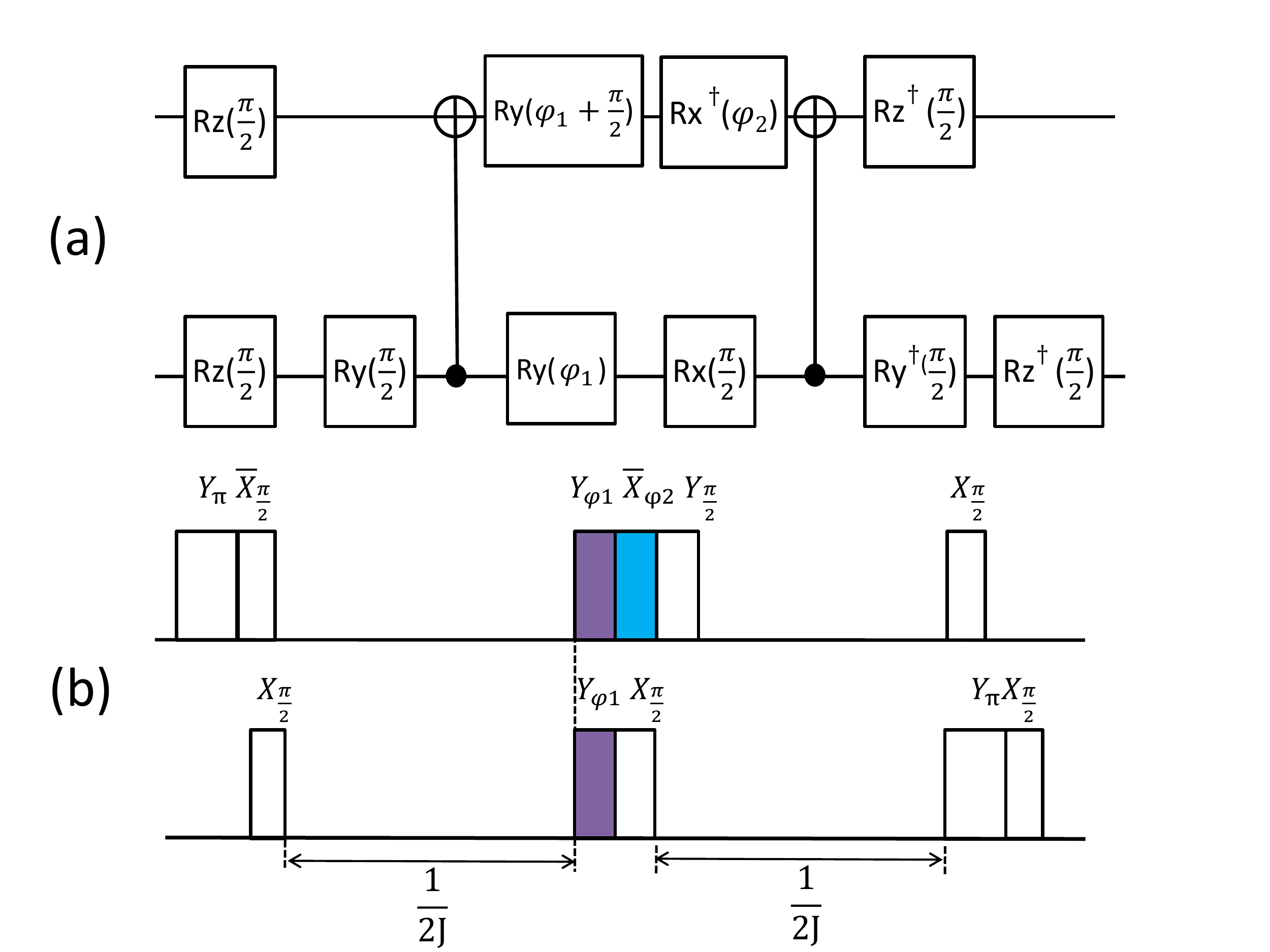} 
			\caption{(a) Quantum circuit for implementing $U_O(B_z)$, where $\varphi_1=\text{arc tan}[(B_z-1)/0]-\pi/4, \varphi_2=|\theta-\pi/2|$, where $\theta=\sqrt{2}B_x/(1-B_z)$. (b) Experimental pulse sequence for implementing $U_O(B_z)$.}\label{supFig33}
		\end{center}
	\end{figure}
	\par It is easy to verify that $U_O({B}_z)\in \textbf{SO}$(4), which can be decomposed as $U_O({B}_z)=\mathcal{M}(A\otimes B)\mathcal{M^{\dagger}}$ \cite{dec2} with $A,B\in \textbf{SU}(2)$ and 
	\begin{eqnarray}
	\begin{matrix}
	\mathcal{M}=\frac{1}{\sqrt{2}}\left(\begin{array}{cccc}1&i&0&0\\0&0&i&1\\0&0&i&-1\\1&-i&0&0\\\end{array}\right).
	\end{matrix}
	\end{eqnarray}
 Any operator in $\textbf{SU}(2)$ can be realized with three rotations via the Euler decomposition, we can thus write $A=R_x(\alpha^A)R_y(\beta^A)R_x(\gamma^A)$ and $B=R_x(\alpha^B)R_y(\beta^B)R_x(\gamma^B)$ \cite{nielsen2002}. Then
	\begin{eqnarray}
	U_O({B}_z) =\mathcal{M}[R_x(\alpha^A)R_y(\beta^A)R_x(\gamma^A)\otimes R_x(\alpha^B)R_y(\beta^B)R_x(\gamma^B)]{\mathcal{M}^{\dagger}}.
	\end{eqnarray}
	$\mathcal M$ can be decomposed as $\mathcal{M}=U_{CNOT}[U_S\otimes (U_HU_S)]$ with 
	$\begin{matrix}
	U_H=\frac{1}{\sqrt{2}}\left(\begin{array}{cc}1&1\\1&-1\\\end{array}\right),
	U_S=\frac{1}{\sqrt{2}}\left(\begin{array}{cc}1&0\\0&i\\\end{array}\right)
	\end{matrix}$ and
	$U_{CNOT}$ as the CNOT gate,
	\begin{eqnarray}
	U_{CNOT}&=&\left(\begin{array}{cccc}1&0&0&0\\0&1&0&0\\0&0&0&1\\0&0&1&0\\\end{array}\right)=\sqrt{i}R_z^1(\pi/2)R_z^2(-\pi/2)R_x^2(\pi/2)exp(-i\frac{\pi}{4} \sigma_z^1\sigma_z^2)R_y^2(\pi/2),
	\end{eqnarray}
	here $R_{\alpha}^j(\beta)=e^{-i\frac{\beta}{2}\sigma_{\alpha}^j}$ denotes a rotation of the $j$-th spin around the axis $\alpha\in \{x,y,z\}$ with $\beta$-angle \cite{CNOT1997}. The quantum circuit to implement $U_O({B}_z)$ is shown in Fig. \ref{supFig33}(a) with the experimental pulse sequence shown in Fig. \ref{supFig33}(b).

	\section{Error analysis}\label{th_error}
	\par The QFI of the final state, $|\widetilde{g}^{\text{exp}}(B_z)\rangle$, equals to the classical Fisher information (CFI) under the optimal measurement, $\{ |v_{Opt}^{m}(\hat {B}_z)\rangle|m=1,2 \}$ \cite{helstrom1969}, which can be computed as
	\begin{equation}\label{eq:FC}
	F^{opt}_C(B_z,\hat{B}_z)= \frac{ [\partial_{B_z}p_1(B_z,\hat{B}_z) ]^2}{p_1(B_z,\hat{B}_z)} + \frac{ [\partial_{B_z}p_2(B_z,\hat{B}_z) ]^2}{p_2(B_z,\hat{B}_z)},
	\end{equation}
where $B_z$ is the parameter to be estimated and $\hat{B}_z$ is the estimated value obtained from the previous data that is used to design the optimal measurement. The probabilities of the measurement results are 
\begin{equation}
p_m(B_z,\hat{B}_z) = | \langle v_{{Opt}}^{m}(\hat {B}_z) \vert \widetilde{g}^{\text{exp}}(B_{z}) \rangle |^2 =  | \langle v_{{loc}}^{m} | U_O(\hat B_{z})\vert \widetilde{g}^{\text{exp}}(B_{z}) \rangle |^2, 
\end{equation}
$m = 1,2$. 
In the experiment, $\partial_{B_z}p_m$ is obtained by the difference method as 
	\begin{equation}\label{eq:Taylor}
	\partial_{B_z}p_1(B_z,\hat{B}_z)|_{\hat{B}_z=B_z}\approx\frac{p_1({B}_z+\delta,,\hat{B}_z)-p_1({B}_z-\delta,\hat{B}_z))}{2\delta},
	\end{equation} 
here $\delta$ is a small value. For simplicity, in the following we will use $p_1(B_z)$ and $p_1(B_z\pm\delta)$ to denote $p_1(B_z,\hat{B}_z)$ and $p_1(B_z\pm\delta,\hat{B}_z)$ respectively. The QFI is then given by
\begin{equation}\label{eq:FQ}
F_Q(B_z)\approx\left(\frac{p_1(B_z+\delta)-p_1(B_z-\delta)}{2\delta}\right)^2\left(\frac{1}{p_1(B_z)}+\frac{1}{1-p_1(B_z)}\right).
\end{equation}
Several errors can affect the accuracy of the QFI: 1) the accuracy of Eq. (\ref{eq:FQ}) depends on the choice of $\delta$; 2) the final state can deviate from the ideal ground state;3) imperfections in the experiment. We give an analysis on how the QFI is affected by these errors. 

\par We first analyze how to choose a reasonable $\delta$ to minimize the error of Eq. (\ref{eq:FQ}). For this purpose we first assume the state is generated ideally, which is denoted as $|\widetilde{g} (B_z)\rangle$, and $p_m^{ideal} (B_z)$, $m=1,2$ are the measurement probabilities on the ideal ground state. We then quantify the difference between the actual partial differentiation and the approximation in Eq. (\ref{eq:FQ}) as 
	\begin{equation}\label{eq:approximation}
	|\partial_{B_z}p_1^{ideal} (B_z)-\frac{p_1^{ideal} (B_z+\delta)-p_1^{ideal} (B_z-\delta)}{2\delta}|.
	\end{equation}
The error on $F_Q$ caused by this difference is 
\begin{equation}\label{eq:Delta1}
\Delta_1=\left(\left(\frac{p_1^{ideal} (B_z+\delta)-p_1^{ideal} (B_z-\delta)}{2\delta}\right)^2-(\partial_{B_z}p_1^{ideal} (B_z))^2\right)\times\left(\frac{1}{p_1^{ideal}(B_z)}+\frac{1}{1-p_1^{ideal}(B_z)}\right),
\end{equation}
here
\begin{equation}
p_1^{ideal} (B_z) =|\langle {v}_{Opt}^1(\hat{B}_z) |\widetilde{g}(B_z) \rangle|^2=\frac{1}{2}\sin(\hat{\theta}-\theta)+\frac{1}{2},
\end{equation} 
with $\tan\hat{\theta}=\frac{\sqrt2 B_x}{1-\hat{B}_z}$ and $\tan\theta=\frac{\sqrt2 B_x}{1-B_z}$. By expanding
$p_1^{ideal} (B_z\pm\delta)$ as 
	\begin{equation}
	p_1^{ideal} (B_z\pm\delta)=p_1^{ideal}(B_z)\pm\delta\frac{\partial p_1^{ideal}(B_z)}{\partial B_z}+\frac{\delta^2}{2!}\frac{\partial^2 p_1^{ideal}(B_z)}{\partial B_z^2}\pm\frac{\delta^3}{3!}\frac{\partial^3 p_1^{ideal}(B_z)}{\partial B_z^3}+o(\delta^4),
	\end{equation}
we get
	\begin{equation}\label{Delta1}
	 \Delta_1=4\delta^2\frac{B_x^2[B_x^2-(B_z-1)^2]}{[(1-B_z)^2+2B_x^2]^4}+O(\delta^4),
	\end{equation}
which reaches its maximum at $B_z=1$ with    
$$
\Delta_1^{\max} \approx\frac{\delta^2}{4B_x^4}.
$$ 
	 
	\par Next we consider the deviation of the final state from the ground state, which can arise from the imperfect adiabatic path and the imperfect experimental operations. We first consider the error induced by the excitation of the imperfect adiabatic path. 
	Without loss of generality, we write the state obtained from the simulated adiabatic path as 
		$\widetilde{\rho}^{sim}(B_z)
		=\sum_{i=1}^4P_i|\widetilde{e}_i(B_z)\rangle\langle\widetilde{e}_i(B_z)|+\sum_{i,j(i\neq j)}(e^{-i\phi_{ij}}P_{ij}|\widetilde{e}_i(B_z)\rangle\langle\widetilde{e}_j(B_z)|+e^{i\phi_{ij}}P_{ij}|\widetilde{e}_j(B_z)\rangle\langle\widetilde{e}_i(B_z)|),$
		where $\{P_i\}$ are the diagonal entries, $P_{ij}$ and $\phi_{ij}$ are the norm and phase of off-diagonal entries of the density matrix in the energy eigen-basis
		 \begin{eqnarray}\label{eigens}
			|\widetilde{e}_1(B_z)\rangle&\equiv&|\widetilde{g}(B_z)\rangle=-\sin\frac{\theta}{2}|11\rangle+\cos\frac{\theta}{2}\frac{|01\rangle+|10\rangle}{\sqrt2}, \notag\\
			|\widetilde{e}_2(B_z)\rangle&=&\cos\frac{\theta}{2}|11\rangle+\sin\frac{\theta}{2}\frac{|01\rangle+|10\rangle}{\sqrt2}, \notag\\
			|\widetilde{e}_3(B_z)\rangle&=&-\sin\frac{\theta}{2}|00\rangle+\cos\frac{\theta}{2}\frac{|01\rangle-|10\rangle}{\sqrt2}, \notag\\
			|\widetilde{e}_4(B_z)\rangle&=&\cos\frac{\theta}{2}|00\rangle+\sin\frac{\theta}{2}\frac{|01\rangle-|10\rangle}{\sqrt2},
		 \end{eqnarray}
	with $\text{tan}\theta=\frac{\sqrt2B_x}{1-B_z}$. 
	
	The excitations in $\widetilde{\rho}^{sim}(B_z)$ lead to errors in $p_1(B_z)$ and $p_2(B_z)$. 
	Under the optimal measurement, which corresponds to the observable $O_{Opt}(B_z)=|v_{Opt}^1(B_z)\rangle\langle v_{Opt}^1(B_z)|-|v_{Opt}^2(B_z)\rangle\langle v_{Opt}^2(B_z)|$, we have
	 \begin{eqnarray}\label{trrhoM}
		p_1^{sim}(B_z)-p_2^{sim}(B_z)&=&Tr(\widetilde{\rho}^{sim}(B_z)O_{Opt}(B_z))=2P_{12}\cos\phi_{12}.
	 \end{eqnarray}
Together with $p_1^{sim}(B_z)+p_2^{sim}(B_z)=1$, we get $p_1^{sim}(B_z)=\frac{1}{2}+P_{12}\cos\phi_{12}$, while for the ideal ground state, $P_{12}=0$, we have $p_1^{ideal}(B_z)=\frac{1}{2}$. 
	
	
	  
Similarly at $B_z\pm\delta$, we have	
\begin{equation}
p_1^{sim}(B_z\pm\delta)=\frac{1+[P_1(B_z\pm\delta)-P_2(B_z\pm\delta)]\sin(\theta-\theta^{\pm\delta})}{2}+P_{12}(B_z\pm\delta)\cos\phi_{12}(B_z\pm\delta)\cos(\theta-\theta^{\pm\delta})
\end{equation}
where $\theta^{\pm\delta}=\arctan\frac{\sqrt2 B_x}{1-(B_z\pm\delta)}$, and
 $p_1^{ideal}(B_z\pm\delta)=\frac{1+\sin(\theta-\theta^{\pm\delta})}{2}$.
This leads to an error in the QFI as
	\begin{equation}\label{errortrans}
	\aligned
	\Delta_2&\equiv F_Q^{sim}(B_z) -F_Q^{ideal}(B_z)\\
	&\approx\frac{\partial F_Q}{\partial p_1(B_z-\delta)}[p_1^{sim}(B_z-\delta)-p_1^{ideal}(B_z-\delta)]\\
	&+\frac{\partial F_Q}{\partial p_1(B_z)}[p_1^{sim}(B_z)-p_1^{ideal}(B_z)]\\
	&+\frac{\partial F_Q}{\partial p_1(B_z+\delta)}[p_1^{sim}(B_z+\delta)-p_1^{ideal}(B_z+\delta)],
	\endaligned
	\end{equation}
	where 
	\begin{eqnarray}\label{partial}
		\frac{\partial F_Q(B_z)}{\partial p_1(B_z)}&= &\left(\frac{p_1(B_z+\delta)-p_1(B_z-\delta)}{2\delta}\right)^2 \left(-\frac{1}{p_1(B_z)^2}+\frac{1}{(1-p_1(B_z))^2} \right) \notag \\
		\frac{\partial F_Q(B_z)}{\partial p_1(B_z\pm\delta)}  &= & \pm\frac{p_1(B_z+\delta)-p_1(B_z-\delta)}{2\delta^2} \left(\frac{1}{p_1(B_z)}+\frac{1}{1-p_1(B_z)} \right).
	\end{eqnarray}
Note that $\frac{\partial F_Q(B_z)}{\partial p_1(B_z+\delta)}=-\frac{\partial F_Q(B_z)}{\partial p_1(B_z\pm\delta)}$ and $\frac{\partial F_Q(B_z)}{\partial p_1(B_z)}=0$, we then have 
	\begin{eqnarray}\label{Delta2}
	\Delta_2\approx\frac{\partial F_Q(B_z)}{\partial p_1(B_z+\delta)} (\Delta_{sim}^\delta-\Delta_{sim}^{-\delta})=\frac{2\sqrt2B_x}{2B_x^2+(B_z-1)^2} \frac{\Delta_{sim}^\delta-\Delta_{sim}^{-\delta}}{\delta}.
	\end{eqnarray}
	where $\Delta_{sim}^{\pm\delta}=p_1^{sim}(B_z\pm\delta)-p_1^{ideal}(B_z\pm\delta)$.
	
\begin{figure}[htb]
		\begin{center}
			\includegraphics[height=3.5cm,width=1\linewidth]{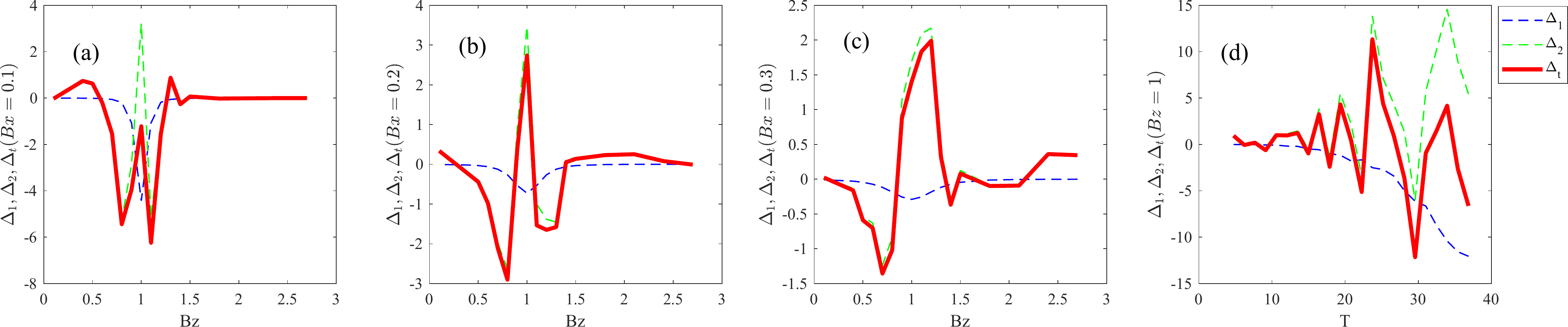} 
			\caption{Numerically simulated $\Delta_{1},\Delta_{2}$ and $\Delta_{t}$ at different $B_z$ for $B_x=0.1,0.2,0.3$ in (a)-(c) and at different $B_x$ for $B_z=1$ in (d), respectively. } \label{supFigErrorAdi}
		\end{center}
	\end{figure}

	
Combined with the error of the approximation, we have
	\begin{eqnarray}
		\Delta_t=\Delta_1+\Delta_2\approx 4\delta^2\frac{B_x^2[B_x^2-(B_z-1)^2]}{[(1-B_z)^2+2B_x^2]^4} +\frac{2\sqrt2B_x}{2B_x^2+(B_z-1)^2} \frac{\Delta_{sim}^\delta-\Delta_{sim}^{-\delta}}{\delta}.
	\end{eqnarray}	
In Fig. \ref{supFigErrorAdi}, we numerically simulate evolution with the decoherence and plot 
$\Delta_{1}$, $\Delta_{2}$ and $\Delta_{t}$, where $\delta$ is taken as $0.03$. It can be seen that $\Delta_{t}$ is in the region of $[-12.2,11.3]$. To quantify the total relative deviation, we take the N experiment points as a N-dimensional vector, denoted as $\vec{F_Q}^{ideal}$ and $\vec{F_Q}^{sim}$, and quantify the total relative deviation as $\frac{\parallel\vec{F_Q}^{dev}\parallel}{\parallel\vec{F_Q}^{ideal}\parallel}$ with $\vec{F_Q}^{dev}=\vec{F_Q}^{sim}-\vec{F_Q}^{ideal}$. 
In the first set of the experiment, at different $B_z$ for $B_x=0.1,0.2,0.3$ the simulation gives a total relative deviation as 18.2\%,  and in the second set of the experiment, for different $B_x$ at $B_z=1$ the simulation gives a total relative deviation as 7.0\% .

\begin{figure}[htb]
		\begin{center}
			\includegraphics[height=9cm,width=1\linewidth]{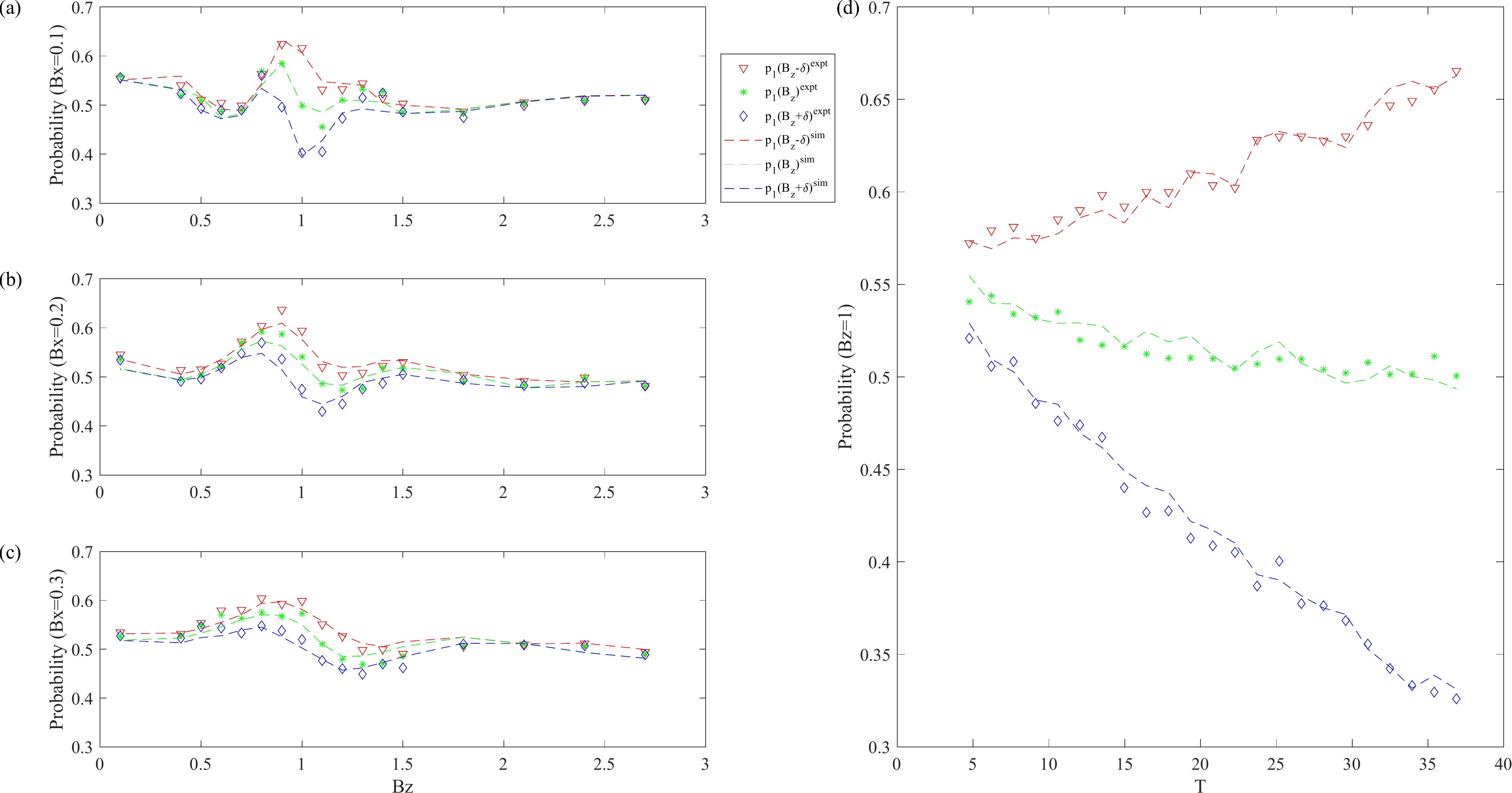} 
			\caption{Experimental measure probabilities of $p_1^{exp}(B_z-\delta),p_1^{exp}(B_z)$ and $p_1^{exp}(B_z+\delta)$ (denoted by the red triangles, green stars and blue diamonds, respectively) for (a)-(c) $B_x=0.1,0.2$ and 0.3, and (d) for the measurement at the critical point $B_z = 1$ with different $B_x$, along with the numerical simulations (denoted by the red, green and blue dashed lines, respectively). These data  correspond to the probabilities to obtain Fig. 3 in main text.}\label{supFig3}
		\end{center}
	\end{figure}

At last we analyze the experimental imperfection, which leads to additional errors in the QFI. Fig. \ref{supFig3} shows the experimental measured probabilities of $p_1^{exp}(B_z+\delta)$, $p_1^{exp}(B_z)$ and $p_1^{exp}(B_z-\delta)$ with $\delta=0.03$, compared with the quantities obtained by the simulation. It can be clearly seen that the experimental results agree well with the numerical simulation. The experimental QFI can be reconstructed as 

\begin{equation}\label{eq:FQexp}
F_Q^{exp}(B_z)\approx\left(\frac{p_1^{exp}(B_z+\delta)-p_1^{exp}(B_z-\delta)}{2\delta}\right)^2 \left(\frac{1}{p_1^{exp}(B_z)}+\frac{1}{1-p_1^{exp}(B_z)}\right).
\end{equation}

And the difference between the experiment and the simulation can be quantified as
\begin{eqnarray}
 \Delta_2^{exp} = F_Q^{exp}(B_z)-F_Q^{sim}(B_z) \approx \frac{2\sqrt2B_x}{2B_x^2+(B_z-1)^2} \frac{\Delta_{exp}^\delta-\Delta_{exp}^{-\delta}}{\delta}
\end{eqnarray}
where $\Delta_{exp}^{\pm\delta} \equiv p_1^{exp}(B_z\pm\delta)-p_1^{sim}(B_z\pm\delta)$.   
The data shows $\Delta_{exp}^\delta-\Delta_{exp}^{-\delta}$ is in the region of $[-0.017, 0.011]$, which leads to  $\Delta_2^{exp}\in [-5.8,6.1]$. Similarly, we can quantify the total relative deviations of $F_Q^{exp}$ from $F_Q^{sim}$ as $\frac{\parallel\vec{F_Q}^{sim}-\vec{F_Q}^{exp}\parallel}{\parallel\vec{F_Q}^{sim}\parallel}$, which is 8.8\% in the first set of the experiment for different $B_z$ at $B_x=0.1,0.2,0.3$ and 5.1\% for the second set of the experiment for different $B_x$ at $B_z=1$. These experimental errors comes from the inhomogeneity of the field, including both the radio frequency field and the static magnetic field, as well as the imperfect calibration of the radio frequency pulses, the noises in the measurement, etc. 

\begin{figure}[htb]
		\begin{center}
			\includegraphics[height=10cm,width=1\linewidth]{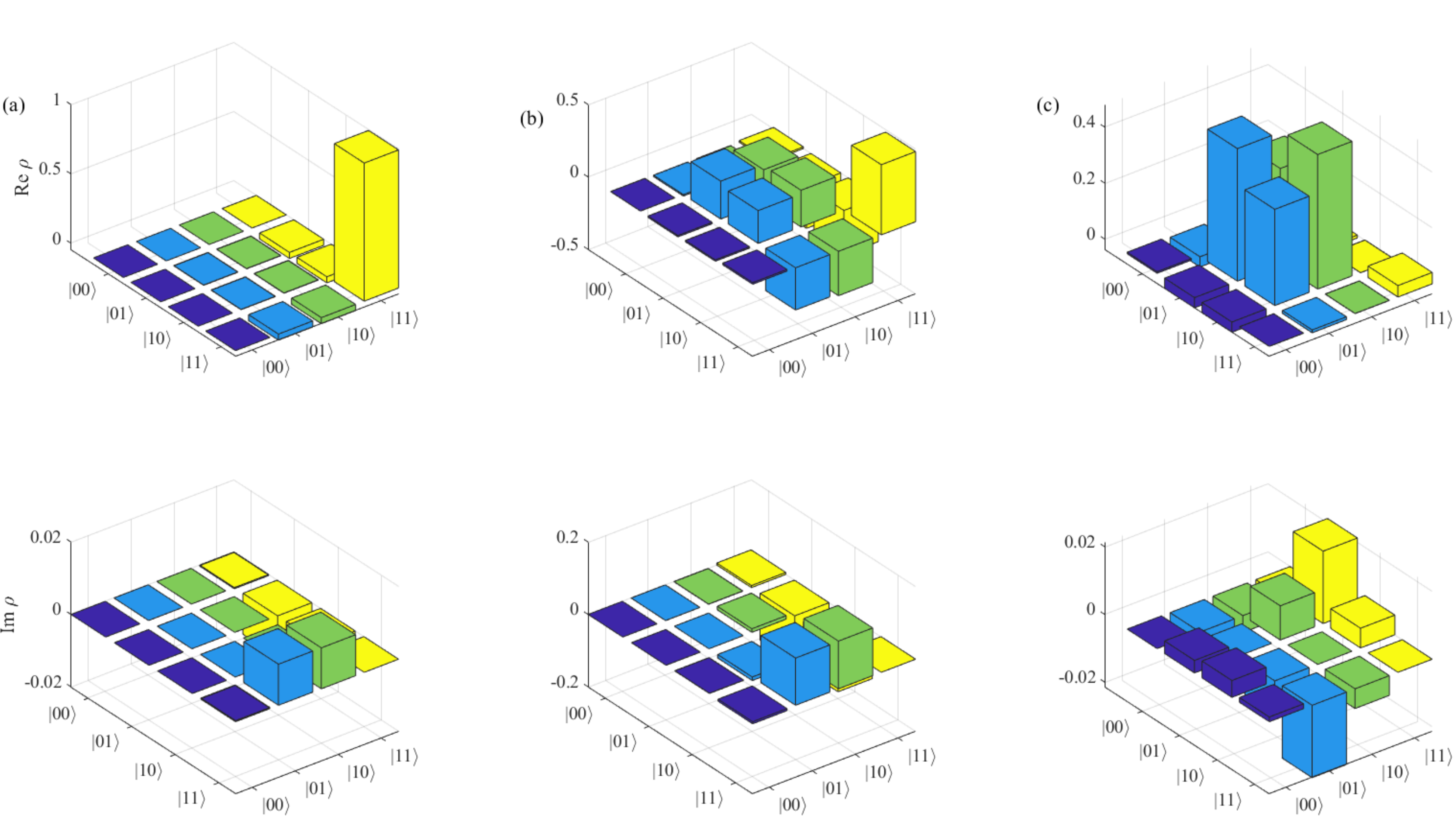} 
			\caption{(a)-(c) correspond to the experimentally reconstructed density matrices of ground states at $B_z=2.7,1$ and 0.1 when $B_x=0.1$, the corresponding fidelities with ideal ground states are $99.6\%, 94.2\%$ and $96.8\%$, respectively.}\label{supFig4}
		\end{center}
	\end{figure}

We also experimentally reconstruct the density matrices \cite{tomoPLA} for the case of $B_x=0.1$, which has the longest evolution time. We show three reconstructed states at $B_z=2.7,1$ and $0.1$ in Fig. \ref{supFig4}. The fidelity between these states and the corresponding ideal ground states are $99.6\%, 94.2\%$ and $96.8\%$ respectively, while the fidelity between these states and the corresponding states obtained via simulation are $99.9\%, 95.7\%$ and $97.6\%$ respectively.

\section{Calibration of the measured effective magnetic field}

The theoretical model  used in the current adiabatic quantum metrology is 
	\begin{eqnarray}
	\label{ham}
	\widetilde{\mathcal{H}}_{\text{Ising}}=B_z(\sigma_{z}^1+\sigma_z^2)+B_x(\sigma_x^1+\sigma^2_x)+\sigma_z^1\sigma_z^2
	\end{eqnarray}
with a small known transverse field $B_x\ll 1$. The unknown magnetic field $B_z$ can be estimated from the measurement probability $p_m(B_z,\hat{B}_z) $ under the optimal projective measurement on $\{ |v_{Opt}^{m}(\hat {B}_z)\rangle|m=1,2 \}$ \cite{helstrom1969} with $p_m(B_{z},\hat{B}_z) = | \langle v_{{Opt}}^{m}(\hat {B}_z) \vert \widetilde{g}(B_{z}) \rangle |^2 $. The measurement uncertainty depends on the previous estimated value $\hat{B}_z$ and the prepared ground state $\vert \widetilde{g}(B_{z}) \rangle$.  The ground state, $\vert \widetilde{g}(B_{z}) \rangle$, has the maximal QFI at the critical point,  $B_{z} = 1$. Consequently, to achieve the best precision, we add a control field $B_c(t)$ along the $z$ axis is to adiabatically drive the system, i.e.,

	\begin{eqnarray}
	\label{ham}
	\widetilde{\mathcal{H}}_{}(t)=[B_z+B_c(t)](\sigma_{z}^1+\sigma_z^2)+B_x(\sigma_x^1+\sigma^2_x)+\sigma_z^1\sigma_z^2. 
	\end{eqnarray}
The unknown magnetic field $B_z$ can be obtained from the adiabatic estimation of $\mathcal{B}_{z}(t) = B_z+B_c(t) $. 

In the experiments, the adiabatic scheme is implemented on the physic system: 

	\begin{eqnarray}
	\label{ham_NMR}
        \mathcal{H}_{NMR} = \frac{\gamma_H B^H_{detuning}}{2} \sigma_{z}^1+ \frac{\gamma_C  B^C_{detuning}}{2} \sigma_z^2+ \frac{ \pi J}{2} \sigma_z^1\sigma_z^2,  
	\end{eqnarray}
where the gyromagnetic ratio $\gamma_H = 267.522 \times 10^6  \text{ rad} \cdot s^{-1} \cdot \text{ Tesla}^{-1}$ and $\gamma_C = 67.283 \times 10^6  \text{ rad} \cdot s^{-1} \cdot \text{ Tesla}^{-1}$. By the double resonance technique, one can set $\gamma_H B^H_{detuning} = \gamma_C  B^C_{detuning} = \mathcal{B}_{z}(t)\pi J $, and the time is also rescaled as $t' = 2t/\pi J $ in the experiments. Here $B^H_{detuning}$ and $B^C_{detuning} $ are the effective magnetic fields experienced on the nuclear spin $^1$H and $^{13}$C, respectively.  When the system is drove to the critical point in order to achieve the maximal QFI, the requirement should be satisfied: 

	\begin{eqnarray}
	\label{B_eff}
        \gamma_H B^H_{detuning} = \gamma_C  B^C_{detuning} = \pi J.   
	\end{eqnarray}
Consequently, the effective magnetic fields $B^H_{detuning}  \approx  2.5 \times 10^{-6} $ Tesla and $B^C_{detuning} \approx 10.0\times 10^{-6}$ Tesla can be measured at the critical point. 

\section{Performance: sensitivity, accuracy and bandwidth}

The following analysis will be focused on the adiabatic estimation of $B_{z}$.

\subsection{Sensitivity} 
 The sensitivity is one most important aspect to be characterized for magnetometers, which determines the precision of the device. In our experiments, there are two main kinds of noise that influence the sensitivity for  the adiabatic estimation of the parameter $B_{z}$. One is the quantum fluctuation induced by Heisenberg uncertainty principle on a fundamental level, and the other is  the classical fluctuation caused by electronic noises in the experiment devices. 
		
In the experiment, $p_m(B_{z},\hat{B}_z)$ is obtained from the measurement corresponding to the optimal observable 
\begin{equation}
	M =|v_{Opt}^1(\hat{B}_z)\rangle\langle v_{Opt}^1(\hat{B}_z)|-|v_{Opt}^2(\hat{B}_z)\rangle\langle v_{Opt}^2(\hat{B}_z)|
\end{equation}
The precision can be characterized via the error-propagation as \cite{errorpro}
	\begin{equation}
	(\Delta B_{z})^2 =\frac{(\Delta M)^2}{\left|\frac{\partial\langle  M \rangle }{\partial B_{z}}\right|^2}, 
	\label{err_prop}
	\end{equation}
where \begin{equation}
	(\Delta M )^2 = \langle  M^2 \rangle -  \langle  M \rangle^2,
	\end{equation}
and	$\langle .\rangle$ denotes the expectation value.
The precision of the estimation depends on how sensitive  $\langle M\rangle$ is to the change of $B_{z}$, and also on the variance of $M$. 

We first consider the quantum fluctuation. For the ground state $| g(B_z) \rangle$, it is straightforward to obtain $\langle M\rangle=\sin(\hat\theta-\theta)$, here $\hat{\theta}$ and $\theta$ are determined by  $\tan\hat{\theta}=\frac{\sqrt2B_x}{1-\hat{B}_z}$ and $\tan\theta=\frac{\sqrt2B_x}{1-B_z}$ respectively, $\left|\frac{\partial\langle M\rangle}{\partial B_z}\right|=\frac{\sqrt2 B_{x}}{\left[\left(1-B_{z}\right)^{2}+2 B_{x}^{2}\right]}$. As the QFI of $| g(B_z) \rangle$ is  $F_Q(B_z)=\frac{2 B^2_{x}}{\left[\left(1-B_{z}\right)^{2}+2 B_{x}^{2}\right]^2}$, we then have $\left|\frac{\partial\langle M\rangle}{\partial B_z}\right|=\sqrt{F_Q(B_z)}$. Therefore, 
\begin{equation}
(\Delta B_{z})_q^2=\frac{1 - \sin^2(\hat\theta-\theta)}{F_Q(B_z)}.
\end{equation}
In our experiment, the spin ensemble is initially in the PPS state as $\rho_g=\frac{1-\epsilon}{4}\hat1+\epsilon|g(B_z)\rangle\langle g(B_z)|$ with the thermal polarization $\epsilon \sim 10^{-5}$ at the room temperature. If the polarization factor is included, we have $(\Delta B_{z})_q^2 |_{\rho_g}= [1 - \epsilon^2 \sin^2(\hat\theta-\theta)]/[\epsilon^2 F_Q(B_z)]$. If the sample contains $N_m$ uncorrelated molecules, the precision due to the quantum fluctuation is then given by
$$
(\Delta B_{z})_q^2 |_{N_m,\rho_g}= \frac{1 - \epsilon^2 \sin^2(\hat\theta-\theta)}{N_m \epsilon^2 F_Q(B_z)}. 
$$

Apart from the quantum fluctuation, the other main factor that affects $(\Delta M )^2$ is the classical noise in the measurement devices, which can be characterized by the Signal to Noise ratio (SNR), i.e., $(\Delta M )_c^2 = 1/SNR^2$. Again, according to the error-propagation formula Eq.\eqref{err_prop}, the precision for the estimation of $B_z$ caused by the classical noise is given by $$
(\Delta B_{z})_c^2 = \frac{1}{SNR^2 F_Q(B_z)}. 
$$
At the presence of both the quantum fluctuation and the classical noise, 
$$
(\Delta B_{z})^2 = \frac{1 - \epsilon^2 \sin^2(\hat\theta-\theta)}{N_m \epsilon^2 F_Q(B_z)} + \frac{1}{SNR^2 F_Q(B_z)}.
$$

In our experiment, $N_m\sim 10^{20}$, $\epsilon \sim 10^{-5}$, the precision caused by the quantum fluctuation is thus $(\Delta B_{z})_q^2 |_{N_m,\rho_g} \sim 10^{-10}/F_Q(B_z)$, while $SNR \sim 10^3$ in the experiment, the precision caused by the classical noise is then $(\Delta B_{z})_c^2 \sim 10^{-6}/F_Q(B_z)$. Therefore, the uncertainty from the classical noise is much larger than the quantum noise. When $B_x=0.1$, at the critical point, $F_Q(B_z=1)=\frac{T^2}{2c^2}$ with $c=2.3$ and $T=\frac{c}{B_x}=23$, we have $\Delta B_z \sim\frac{\sqrt2c}{ 10^3T} : \sim 10^{-6}\text{Hz}^{-1}$. This corresponds to $\sim 10^{-11}\text{ Tesla}\cdot \text{Hz}^{-1}$ if converted to the sensitivity of the effective magnetic field on the $^{13}C$ spin($\gamma_C = 67.283 \times 10^6  \text{ rad} \cdot s^{-1} \cdot \text{ Tesla}^{-1}$).
 	
\subsection{Bandwidth}
	Due to the measured signal $\langle g(B_z)|M(\hat{B_z})|g(B_z)\rangle=\sin(\hat\theta-\theta)$, its response to the encoded parameter $B_z$ thus is
       \begin{equation}
       \mathcal R(B_z)=\left|\frac{\partial\langle g(B_z)|M(\hat{B_z})|g(B_z)\rangle}{\partial B_z}\right|=\left|\cos (\hat\theta-\theta)\frac{\partial \theta}{\partial B_z}\right|\approx\frac{2 B_{x}}{\left(1-B_{z}\right)^{2}+2 B_{x}^{2}}. 
       \end{equation}
     Here we assume that the magnetometer works near $\hat{B_z}$. This shows that the response $\mathcal R(B_z)$ decreases as $B_z$ is away from $\hat{B_z}$.
     
     \noindent Bandwidth is defined as the range over which the response $\mathcal R(B_z)$ degrades by $1/\sqrt2$ times of the maximum, i.e., $\mathcal R(B_z)\ge\frac{\max[R(B_z)]}{\sqrt2}$.  It is easy to obtain the bandwidth as $2\sqrt{2(\sqrt2-1)}B_x$. Taking the case $B_x=0.1$ as an example, the bandwidth is $~0.18$, which corresponds to $1.8\times10^{-6}$ Tesla as an effective magnetic field on $^{13}$C.
	
\subsection{Accuracy}
The accuracy of our adiabatic protocol results from the imperfect ground-state prepared in the experiments.  With the ideal ground state $\widetilde{\rho}^{ide}(B_z)$ under the observable $M(\hat{B}_z)$,   the unknown $B_z$ can be deduced from the measurement result as
\begin{equation}\label{dedu_fun}
	B_z^{deduced}=1+\frac{\sqrt2B_x}{\tan[\hat{\theta}-\arcsin(m)]},
\end{equation}where the measurement result  $m=Tr[\widetilde{\rho}^{ide}(B_z)M(\hat{B}_z)]=\sin(\hat\theta-\theta)$, $\tan\hat{\theta}=\frac{\sqrt2B_x}{1-\hat{B}_z}$ and $\tan\theta=\frac{\sqrt2B_x}{1-B_z}$.  When $\hat{B}_z=B_z$, the measurement is optimal, leading to  $B_z^{deduced}=B_z$ for the ideal ground state.  The discrepancy between the experimentally obtained state $\widetilde{\rho}^{exp}(B_z)$ and the theoretical ground state $\widetilde{\rho}^{ide}(B_z)$ will lead to a deviation of the estimated value $B_z^{deduced}$ from the true value $B_z$: 
		\begin{equation}\label{dev_Bz}
		\delta B_z=B_z^{deduced}-B_z\approx\frac{\partial B_z^{deduced}}{\partial m}\delta m=\frac{\delta m[(1-B_z)^2+2B_x^2]}{\sqrt2B_x}=\frac{\delta m}{\sqrt{F_Q}},
		\end{equation}
		where $\delta m=Tr[\widetilde{\rho}^{exp}M]-Tr[\widetilde{\rho}^{ide}M]=Tr[\widetilde{\rho}^{exp}M]$  caused by the imperfect ground state.  Obviously, for the same deviation $\delta m$, the state with larger QFI can lead to higher accuracy.
		
		Therefore, the accuracy of the adiabatic quantum metrology $\delta B_z$ depends on both the the systematic  deviation of the measurements $\delta m$ caused by the imperfection of ground-state preparation and QFI of the ground state. By reconstructing the experimentally prepared ground states using quantum state tomography \cite{tomoPLA}, we can characterize the accuracy in the experiments. Taking three representative states at $B_z=2.7, 1$ and $0.1$ at $B_x=0.1$ for the illustration, $\delta m = Tr[\widetilde{\rho}^{exp}M] = 0.01, 0.02$
and $0.15$, respectively, resulting in the corresponding accuracy  $\delta B_z =0.180,0.003$ and $44.886$. Converting these results as the effective magnetic field on $^{13}$C, they correspond to $1.8\times10^{-6}, 2.9\times10^{-8}$ and $4.5\times10^{-4}$ T respectively. The results show that the ground state at critical point has the highest accuracy since it has the largest QFI.

\section{Comparison with the standard scheme of quantum metrology}

In the conventional scheme, the standard procedure to estimate the magnitude of a magnetic field with two spins is to prepare a maximally entangled state, $\frac{|00\rangle+|11\rangle}{\sqrt{2}}$, then let it evolve under the Hamiltonian $B_z(\sigma^1_z+\sigma^2_z)$. For the noise-free evolution, the QFI of the standard scheme scales as $T^{2}$, i.e., it can achieve the Heisenberg scaling. For noisy evolution, the probe will lose information due to the interactions with the environment, the QFI typically first increases to its maximum then decays to zero. The general procedure is then to stop the process at its maximum(at some time $t$) and repeat the process $T/t$ times. In this procedure, the QFI scales as $\frac{T}{t}F_Q(t)\le T\left[\frac{F_Q(t)}{t}\right]_{\text{max}}\sim T$, i.e., the shot noise limit \cite{Zhou2018}.


While in the adiabatic quantum metrology, the probe state remains as the ground state of the Hamiltonian. Thus as long as the temperature is small compared to the minimum energy gap of the system \cite{robust_ad}, the state is robust against the decays induced by the interactions with the thermal environment. The robustness of the adiabatic process against the decay has been well studied. Here we simulate the process using the master equation which describes a system of spins coupled to an environment captured by a spectrum $g(\omega)$ \cite{robust_ad},
\begin{eqnarray}\label{master}
	\begin{aligned}
		\frac{\mathrm{d} \rho}{\mathrm{d} t}=&-i\left[H_{S}, \rho\right]-\sum_{i, a, b}\left[N_{b a}\left|g_{b a}\right|^{2}\left\langle a\left|\sigma_{-}^{(i)}\right| b\right\rangle\left\langle b\left|\sigma_{+}^{(i)}\right| a\right\rangle\right.\left.+\left(N_{a b}+1\right)\left|g_{a b}\right|^{2}\left\langle b\left|\sigma_{-}^{(i)}\right| a\right\rangle\left\langle a\left|\sigma_{+}^{(i)}\right| b\right\rangle\right]\\
		&\{(|a\rangle\langle a| \rho)+(\rho|a\rangle\langle a|)-2|b\rangle\langle a|\rho| a\rangle\langle b|\},
		\end{aligned}
\end{eqnarray}
where the states $|a\rangle$ are the instantaneous eigenstates of the time-dependent Hamiltonian, $H_S$, with the eigen-energy $\omega_a$,
\begin{eqnarray}
	N_{b a}=\frac{1}{\exp \left[\beta\left(\omega_{b}-\omega_{a}\right)\right]-1}
\end{eqnarray}
is the Bose-Einstein distribution at temperature $1/\beta$, and
\begin{eqnarray}
	g_{b a}=\left\{\begin{array}{ll}
		\lambda g\left(\omega_{b}-\omega_{a}\right), & \omega_{b}>\omega_{a} \\
		0, & \omega_{b} \leqslant \omega_{a}
		\end{array}\right.
\end{eqnarray}
For simplicity, we choose $g(\omega)=1$ for $\omega\ge0$ and zero otherwise. 
In the simulation, we take $\lambda=0.4$ and simulate the process at temperatures $1/\beta=0.001, 0.01, 0.02, 0.5$, which represent the temperature varies from $1/\beta\ll\Delta_{min}$ to $1/\beta\gg\Delta_{min}$ with the minimal energy gap $\Delta_{min}=2\sqrt2B_x$ as mention in Eq. \eqref{gap}. 
\begin{figure}[htb]
	\begin{center}
		\includegraphics[height=6cm,width=0.4\linewidth]{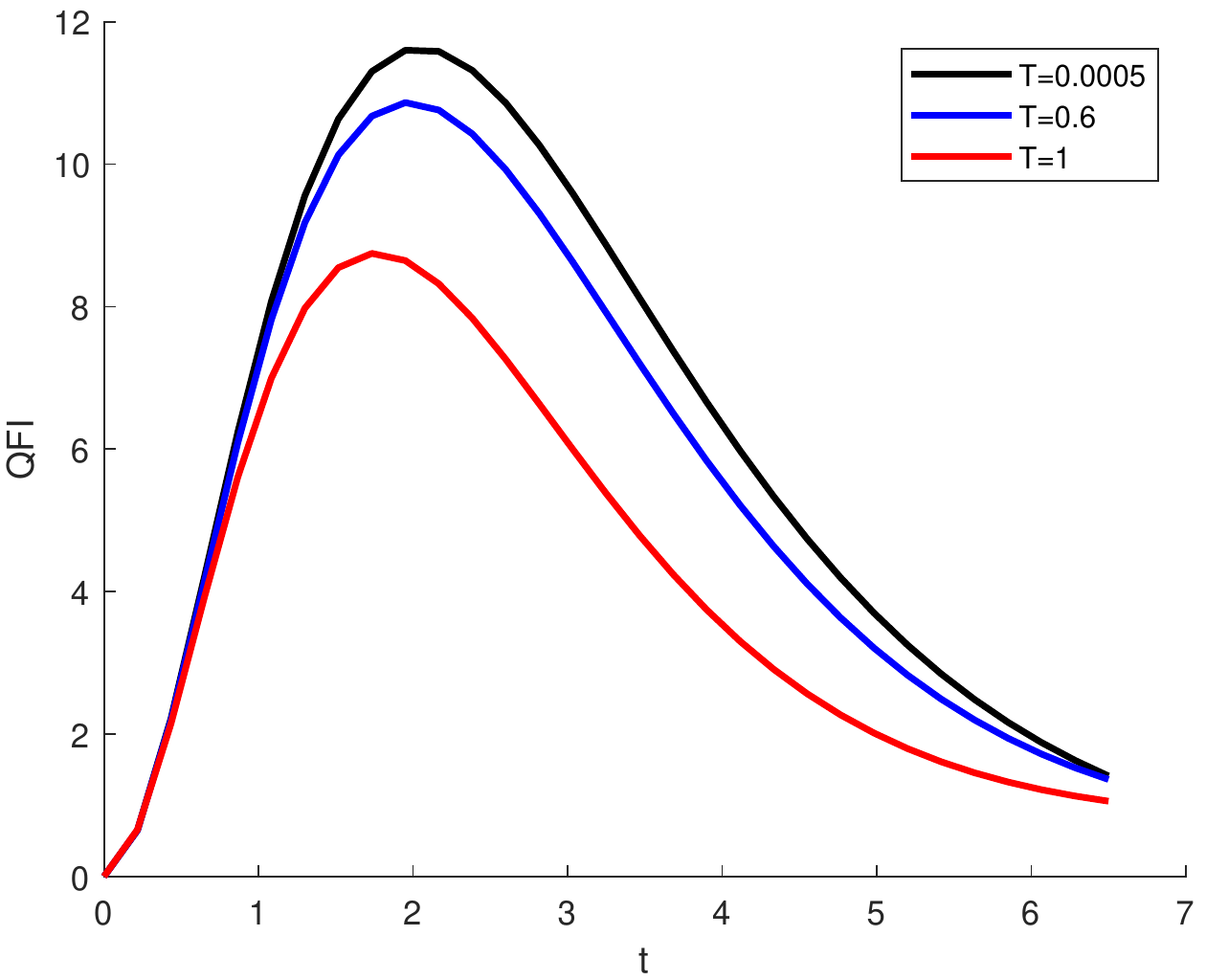}
		\includegraphics[height=6cm,width=0.4\linewidth]{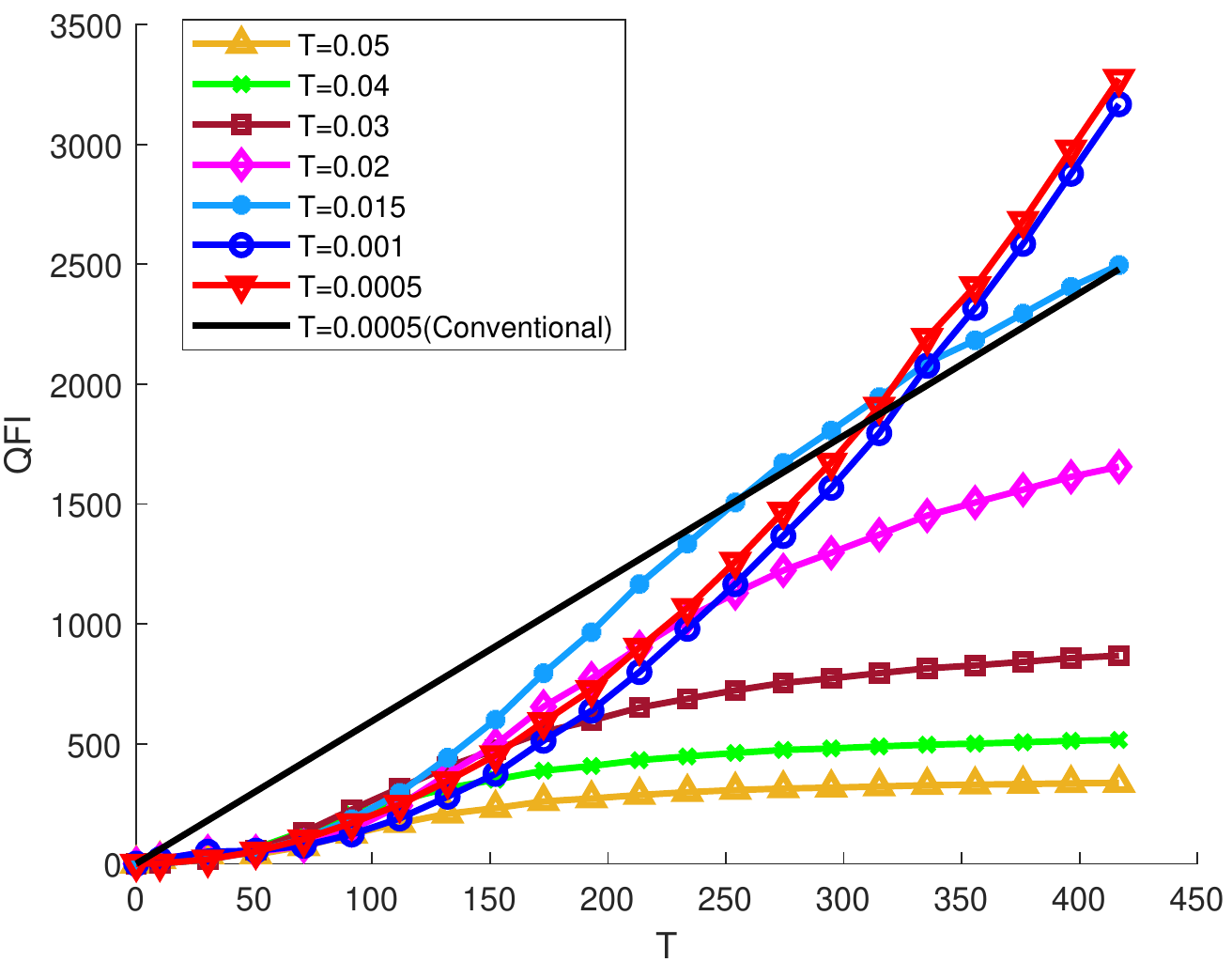}\\
		\caption{Numerical simulation of QFI for (a) the conventional scheme at different temperatures and (b) comparison between the conventional and the adiabatic schemes at different temperatures.}\label{supFig6}
	\end{center}
\end{figure}

The QFI of the standard scheme under the interaction with the environment first increases to its maximum then decays to zero as shown in Fig. \ref{supFig6}(a). Thus for $T\gg 1/\lambda$, the maximal QFI can be obtained is $\frac{T}{t_{max}}F_Q(t_{max})\sim T$, which is the shot noise limit. For the adiabatic scheme, as shown in Fig. \ref{supFig6}(b), when the temperature $1/\beta\ll\Delta_{min}$, it is robust against the decay and the QFI can maintain the Heisenberg scaling.  

\bigskip

	\bibliographystyle{apsrev4-2}
	\bibliography{sup_reference}